\input amstex
\input psfig.sty
\magnification 1200
\TagsOnRight
\def\qed{\ifhmode\unskip\nobreak\fi\ifmmode\ifinner
\else\hskip5pt\fi\fi\hbox{\hskip5pt\vrule width4pt height6pt
depth1.5pt\hskip1pt}}
\define \ds{\displaystyle}
\define \bR {\bold R}
\def\stretch {\noalign{\medskip}}
\define \bC {\bold C}
\define \bCp {\bold C^+}
\define \bCm {\bold C^-}
\define \bCpb {\overline{\bold C^+}}
\define \bCmb {\overline{\bold C^-}}
\define \bm {\bmatrix}
\define \endbm {\endbmatrix}
\baselineskip 18 pt
\parskip 4 pt

\centerline {\bf DETERMINING THE SHAPE OF A HUMAN VOCAL TRACT}
\centerline {\bf FROM PRESSURE MEASUREMENTS AT THE LIPS}

\vskip 5 pt

\centerline {Tuncay Aktosun}
\vskip -8 pt
\centerline{Department of Mathematics}
\vskip -8 pt
\centerline {University of Texas at Arlington}
\vskip -8 pt
\centerline{Arlington, TX 76019-0408, USA}

 \centerline {Alicia Machuca}
\vskip -8 pt
\centerline{Department of Mathematics and Computer Science}
\vskip -8 pt
\centerline {Texas Woman's University}
\vskip -8 pt
\centerline{Denton, TX 76204, USA}

\centerline {Paul Sacks}
\vskip -8 pt
\centerline {Department of Mathematics}
\vskip -8 pt
\centerline {Iowa State University}
\vskip -8 pt
\centerline {Ames, IA 24061, USA}

\vskip 10 pt

\noindent {\bf Abstract}: The inverse problem of determining the cross-sectional
area of a human vocal tract during the utterance of a vowel is considered in terms
of the data consisting of the absolute value of sound pressure at the lips. If the upper
lip is curved downward during the utterance, it is shown that
there may be up to an $M$-fold nonuniqueness
in the determination, where
$M$ is the maximal number of eligible resonances associated with
a related Schr\"odinger operator.
Each of the $M$ such distinct candidates for the vocal-tract area
corresponding to the same absolute pressure is uniquely determined. The
mathematical theory is presented for the recovery of each candidate
for the vocal-tract area, and the admissibility criterion for each of the
$M$ candidates to be a vocal-tract radius
is specified. On the other hand, if the upper lip is horizontal or curved
upward during the utterance, then the inverse problem has a unique solution.
The theory developed is illustrated with some examples.

\vskip 10 pt
\noindent {\bf PACS (2010):}  02.30.Zz, 43.72.Ct\hfil
\vskip -8 pt
\par \noindent {\bf Mathematics Subject Classification (2010):}
34A55, 35R30, 76Q05
\vskip -8 pt
\par\noindent {\bf Keywords:}
Inverse scattering, Speech acoustics, Shape of vocal tract
\vskip -8 pt
\par\noindent {\bf Short title:}
Determining the shape of a vocal tract

\newpage

\noindent {\bf 1. INTRODUCTION}
\vskip 3 pt

The production of human speech is similar [7,8,13,15-17] in principle to the
sound production in tubular musical instruments.
The inhaled air travels from the mouth
down the vocal tract, a tube about 14-20 cm in length with
a pair of lips at the mouth and another pair
of lips known as the vocal cords at the opposite end.
The inhaled air
passes between the vocal cords and travels
to the lungs through another tube known as
the voice box or the larynx. From the lungs the air travels back and
enters the vocal tract through the glottis
(the opening between the vocal cords).
The pressure created
at the glottis puts the nearby air molecules in the vocal tract into longitudinal vibrations.
These vibrations are responsible for the propagation of the sound
pressure in the vocal tract. The pressure
wave comes out of the mouth and is transmitted in the air to a person's ears or to a
microphone.

Human speech consists of basic units known as phonemes. For example,
when we utter the word ``book" we in succession produce
the three phonemes
/b/, /u/, and /k/, each lasting
about 0.1 sec.
The phonemes
can be classified into two main groups as vowels and consonants.
For the
mathematical description of
vowel production,
one can satisfactorily ignore the articulators (such as the tongue) and
assume that the cross-sectional area $A(x)$
of the vocal tract as a function of the distance from the glottis is the
only factor responsible for the produced sound.
We mention that $x=0$ corresponds to the glottis and $x=\ell$ to the lips
located at $x=\ell,$ where $\ell$ denotes
the length of the vocal tract.

During the production of
each vowel, the shape of the vocal tract can be assumed
not to depend on time.
In some sense this is analogous to watching a movie, where each frame contains
a static image and lasts a short period. The continuous movie is perceived
and the continuous speech is
heard, respectively, as we watch a succession of static images and as we encounter a succession
of static shapes of the vocal tract.
The length $\ell$ of the vocal
tract depends on the individual speaker, and it can be satisfactorily
assumed that $\ell$ does not change
during the phoneme production. Mathematically,
one can ignore the bending of the vocal tract
and assume that the cross-sectional area at each $x$-value
is circular with radius $r(x),$ where
$$A(x)=\pi[r(x)]^2.\tag 1.1$$

Let us use $p(x,t)$ and $v(x,t)$ for
the pressure and the volume velocity, respectively, in the vocal tract at location $x$ and
at time $t.$
Let $\nu$ denote the frequency measured in Hz (Hertz),
$\omega$ the angular frequency in rad/sec, $k$ the angular
wavenumber in rad/cm, and $\lambda$ the wavelength in cm.
These quantities are related to each other as
$$\omega=kc,\quad \omega=2\pi \nu,\quad \nu\lambda=c,\tag 1.2$$
where $c$ is the sound speed in the vocal tract, which .
can be assumed to have the same constant value throughout the 
vocal tract. From (1.2) we see that
$$k=\ds\frac{2\pi \nu}{c},$$
and hence $k$ can be viewed as a constant multiple of the frequency.
The audible frequencies usually range from 20 Hz
to 20000 Hz for human beings. So, 20 Hz corresponds to $k=0.0037$ rad/cm
and 20000 Hz corresponds to $k=3.7$ rad/cm.
There is one more relevant constant,
namely the air density $\mu$ used in the
description of the sound propagation in the vocal tract.

If the propagating pressure wave $p(x,t)$ is monochromatic, it contains only
one sinusoidal component at a single frequency. A similar
remark also applies to the volume velocity $v(x,t).$
In general, each of $p(x,t)$ and $v(x,t)$ is a linear combination of
sinusoidal components at many frequencies, and they can be expressed as
$$p(x,t)=\ds\frac{1}{2\pi}\int_{-\infty}^\infty dk\, P(k,x)\,e^{i\omega t},\quad v(x,t)=\ds\frac{1}{2\pi}\int_{-\infty}^\infty dk\, V(k,x)\,e^{i\omega t},\tag 1.3$$
where we refer to $P(k,x)$ and $V(k,x)$ as the pressure and the volume velocity,
respectively, in the frequency domain.
The use of the complex exponent $e^{i\omega t}$
in (1.3) is mathematically convenient. One can certainly avoid using
negative frequencies in (1.3) by noting [1,2] that
$$P(-k,x)=P(k,x)^\ast,\quad V(-k,x)=V(k,x)^\ast,\qquad k\in\bR,\tag 1.4$$
where $\bR:=(-\infty,+\infty)$ and the asterisk denotes complex conjugation.

The time factor $e^{i\omega t}$ in (1.3) is the acoustician's convention, whereas
it is customary to use the time factor $e^{-i\omega t}$
as the physicist's convention.
In this paper we use the former convention, and hence we visualize $e^{ikx}$
in the frequency domain
as the wave component $e^{ikx+i\omega t}$ in the time-domain signal, which is
a plane wave moving in the negative $x$-direction. Similarly,
$e^{-ikx}$ can be visualized as a plane wave moving in the
positive $x$-direction.

Throughout our paper we assume that the vocal-tract radius belongs to class $\Cal A$ specified
below.

\noindent {\bf Definition 1.1} {\it The vocal-tract radius $r(x)$ belongs to class $\Cal A$
if the following conditions are satisfied:}

\item{(a)} {\it The function $r(x)$ is real valued and positive on $x\in(0,\ell).$}

\item{(b)} {\it The function $r(x)$ has positive limits as $x\to 0^+$ and as $x\to \ell^-.$}

\item{(c)} {\it The derivative
function $r'(x)$ is continuous for $x\in(0,\ell)$ and has
finite limits as $x\to 0^+$ and as $x\to \ell^-.$}

\item{(d)} {\it The second-derivative
function $r''(x)$ is integrable on $x\in(0,\ell).$}

\noindent Without any loss of generality, we let
$$r(0):=r(0^+),\quad r'(0):=r'(0^+), \quad r(\ell):=r(\ell^-),\quad r'(\ell):=r'(\ell^-).$$

Our primary goal in this paper is the analysis of the inverse problem of
recovery of the vocal-tract shape when we know the absolute sound
pressure at the lips at all positive frequencies during the production of a vowel.
Mathematically, this is equivalent to the recovery of
$A(x)$ for $x\in(0,\ell)$
when our input data set consists of
 $|P(k,\ell)|$ for $k>0.$ In the analysis of our inverse
 problem we assume that the value
 of $\ell$ is known.
At the end of Section~4 we comment on the solution of the inverse problem
if the value of $\ell$ is not known.
 With the help of (3.33), (3.34), and (3.36), we can argue
 that the input data set in the audible frequency
 range is sufficient for a satisfactory
 approximate recovery. Because of (1.1), we
can equivalently recover $r(x)$ instead of $A(x).$
Furthermore, with the help of (1.4) we see
that our input data set is equivalent to
the data set $|P(k,\ell)|$ for $k\in\bR.$

The associated direct problem can be described as
the determination of the absolute pressure at the lips when the
vocal-tract area function is known. A solution to this direct problem
is explicitly given [1,2] with the help of the Jost function and the Jost
solution associated with a corresponding selfadjoint Schr\"odinger operator on the half line.
A solution to the inverse problem has been given in [1,2]
under a certain additional restriction on $r(x)$
that is equivalent to the absence of
bound states in the associated Schr\"odinger operator.
In our current paper, we study our inverse problem without
that restriction. We
show that the corresponding Schr\"odinger
operator does not have any bound states for certain vowels but it
 has exactly one bound state for the remaining vowels. We provide
a characterization of the absence or presence of a bound state
in terms of the bending of the vocal-tract radius at the lips, i.e.
in terms of the sign of $r'(\ell).$ From the appearance of the lips during
a vowel utterance, one may tell what the sign of $r'(\ell)$ is. For example, for
the vowels /u/, /a/, and /o/, we have $r'(\ell)=0,$ $r'(\ell)>0,$ and
$r'(\ell)<0,$ respectively. We show that
there are no bound states
if $r'(\ell)\ge 0$ and that there is exactly one bound state if
$r'(\ell)<0.$ We provide the solution to our inverse problem in the possible presence of a bound
state, and we also characterize the uniqueness or nonuniqueness that may be occurring
in the presence of a bound state.

Our paper is organized as follows. In Section~2 we present
the preliminaries needed for later sections, by providing
the solution to the direct problem in two different ways.
In Theorem~2.1  we provide the explicit expressions for the pressure and volume
velocity in the vocal tract
in terms of the unique solution $f(k,x)$ to the initial-value
problem consisting of (2.10) and (2.14)
as well as the quantity $F(k)$ appearing in (2.18).
In Theorem~2.2 we
indicate that the solution to the direct problem could alternatively be obtained by solving
the initial-value problem related to (2.2), (2.24), and (2.26). The result in Theorem~2.1
is needed for the formulation of our inverse problem.
The results presented in the
finite interval $x\in(0,\ell)$ in Section~2 are extended
to the half line in Section~3, and this
allows us to exploit the properties
of the absolute pressure at the lips in terms of
the absolute value of the Jost function,
based on the important relationship stated in (2.29).
In Section~3, we elaborate on that relationship and
show that if $r'(\ell)\ge 0$ then the associated Jost function does not
have any bound-state zeros and also show that if $r'(\ell)<0$ then
the Jost function has exactly one bound-state zero. In Section~4
we provide the solution to our inverse problem, with the help
of the Gel'fand-Levitan method and alternatively with the
help of the Marchenko method. We show that
the vocal-tract radius $r(x)$ can uniquely be determined
from the absolute pressure if $r'(\ell)\ge 0.$
On the other hand, if $r'(\ell)<0,$ we show that,
corresponding to
the same absolute pressure,
there are $M$ candidates for $r(x)$ with $r'(\ell)<0$
in addition to the candidate
for $r(x)$ with $r'(\ell)\ge 0.$
We show that
$M$ is the maximal number of eligible resonances [4]
for an associated Schr\"odinger equation and is
uniquely determined by the absolute pressure. In Section~4 we also
prove that each one of the $(M+1)$ candidates for
$r(x)$ corresponding to the same absolute pressure
is uniquely determined. Clearly, only those candidates for which
the constructed $r(x)$ satisfying $r(\ell)>0$ are admissible
as the vocal-tract radii, and we present an
equivalent admissibility criterion.
In Section~5, under the assumption that $r'(\ell)\ge 0,$
we present a time-domain method to solve our inverse  problem,
providing an alternate method to the two frequency-domain
methods presented in Section~4. Finally, in Section~6
we illustrate our results with some examples. In particular,
we use the vocal-tract area data from [18] and compute the
corresponding absolute pressure at the lips by using the alternate method
based on Theorem~2.2. We then use that constructed pressure data
as input to the time-domain method described in Section~5 and
determine the corresponding vocal-tract cross-sectional area.
The input area from [18] and the computed area are shown
in the first plot in Figure~6.1, indicating a fairly accurate
numerical solution to our inverse problem.

\vskip 10 pt
\noindent {\bf 2. THE SOLUTION TO THE DIRECT PROBLEM}
\vskip 3 pt

In this section we review the acoustics in the vocal tract, and we associate
it with the Schr\"odinger equation given in (2.10) for $x\in(0,\ell),$
where the potential $q(x)$
is related to the vocal-tract radius function $r(x)$ as in (2.11). In (2.21) and (2.22), respectively, we provide explicit
expressions for the pressure $P(k,x)$
and the volume velocity $V(k,x)$
in terms of $r(x)$
and the key quantity $F(k)$ given in (2.18).
By expressing the absolute pressure at the lips
explicitly as in (2.23), we provide a solution to the
direct problem of recovery of the absolute pressure at the
lips from the vocal-tract radius function. Via Theorem~2.2
we show that the solution of the
direct problem can alternatively be achieved by solving the system (2.2)
with the initial conditions as in (2.24) and (2.26). Even though the latter method is more
straightforward than the former method, the former is needed in the formulation of
the inverse problem.

It is reasonable [15,16] to assume that the sound
propagation in the vocal tract is lossless and planar
and that the acoustics
in the vocal tract is governed [7,8,15-17] by
$$\cases A(x)\,p_x(x,t)+\mu\,v_t(x,t)=0,\\
\noalign{\medskip} A(x)\,p_t(x,t)+c^2\mu\,
v_x(x,t)=0,\endcases\tag 2.1$$
where $t>0$ and $x\in(0,\ell),$
and the subscripts denote the appropriate partial derivatives.
Using (1.1) and (1.3) in (2.1) we obtain
$$\cases \pi r(x)^2\,P'(k,x)+ic\mu k\,V(k,x)=0,\\
\noalign{\medskip} c\mu\,V'(k,x)+i\pi k\,r(x)^2\,P(k,x)=0,\endcases\tag 2.2$$
where the prime denotes the-$x$
derivative. Eliminating $V(k,x)$ in (2.2), we get the Webster horn equation
$$[r(x)^2\,P'(k,x)]'+k^2\,r(x)^2\,P(k,x)=0,\qquad x\in(0,\ell),\tag 2.3$$
or eliminating $P(k,x)$ in (2.2) we get
$$\left[\ds\frac{V'(k,x)}{r(x)^2}\right]'+k^2
\,\ds\frac{V(k,x)}{r(x)^2}=0,\qquad x\in(0,\ell).\tag 2.4$$

In order to solve (2.2) uniquely, we need
two side conditions, which we can choose by specifying
the glottal volume velocity $v(0,t)$ and by assuming that the pressure
wave at the lips goes only out of the mouth, not into the mouth.
One particular choice for $v(0,t)$ is given by
$$v(0,t)=\delta(t),\tag 2.5$$
where $\delta(t)$ is the Dirac delta distribution. Comparing with
the second equality in (1.3) we see that (2.5) is equivalent to
$$V(k,0)=1,\tag 2.6$$
which corresponds to a unit-amplitude, sinusoidal volume velocity at
the glottis at any angular wavenumber $k.$ With the help
of the first line of (2.2), we see that (2.6) is equivalent to
$$P'(k,0)=-\ds\frac{ic\mu k}{\pi\,  r(0)^2}.\tag 2.7$$

The second side
condition is equivalent to
rejecting the wave component proportional to
$e^{ik\ell+ikct}$ and accepting only the wave component proportional to $e^{-ik\ell+ikct}$
in the expression for $p(x,t)$ when $x=\ell.$ It is known [1,2] that such a condition
is equivalent to
$$P'(k,\ell)=-\left[ik+\ds\frac{r'(\ell)}{r(\ell)}\right]
P(k,\ell).\tag 2.8$$
With the help of (2.2), we see that (2.8) is equivalent to
$$k^2V(k,\ell)=\left[ik+\ds\frac{r'(\ell)}{r(\ell)}\right]V'(k,\ell).\tag 2.9$$

By letting
$$\psi(k,x):=r(x)\,P(k,x),$$
we can transform (2.3) into the Schr\"odinger equation
$$\psi''(k,x)+k^2\psi(k,x)=q(x)\,\psi(k,x),
\qquad x\in(0,\ell),\tag 2.10$$
where the potential $q(x)$ is related to the vocal-tract radius $r(x)$ via
$$q(x):=\displaystyle\frac{r''(x)}{r(x)},
\qquad x\in(0,\ell).\tag 2.11$$
When $r(x)$ belongs to class $\Cal A$ specified in Definition~1.1,
the potential $q(x)$ is real valued and integrable on $x\in(0,\ell).$
In order to analyze the direct and inverse problems associated with
(2.10), it is convenient to supplement (2.10) with
the boundary condition [5,9,11,12]
$$\psi'(k,0)+(\cot\theta)\,\psi(k,0)=0,\tag 2.12$$
where the boundary parameter $\cot\theta$ is related to the
vocal-tract radius as
$$\cot\theta=-\displaystyle\frac{r'(0)}{r(0)}.\tag 2.13$$
We remark that the parameter $\theta$ in (2.12) is allowed to take
any value in the interval $(0,\pi)$ and hence $\cot\theta$ can
be any number in the interval $(-\infty,+\infty).$

Two particular
solutions to (2.10) are relevant for (2.3).
One of these is the solution $f(k,x)$
satisfying the initial conditions
$$f(k,\ell)=e^{ik\ell},\quad f'(k,\ell)=ik\,e^{ik\ell}.\tag 2.14$$
Since $k$ appears as $k^2$ in (2.10), the quantity $f(-k,x)$ is also
a solution to (2.10), and it follows from (2.14) that, for each fixed real
 nonzero $k,$ the functions
$f(k,x)$ and $f(-k,x)$ are linearly independent.
In Section~3 we will relate $f(k,x)$ to the
Jost solution to the half-line Schr\"odinger equation
(3.1) with the asymptotics
$e^{ikx}[1+o(1)]$ as $x\to+\infty.$

The second relevant particular solution to (2.10)
is the solution $\varphi(k,x)$ satisfying the
initial conditions
$$\varphi(k,0)=1,\quad \varphi'(k,0)=-\cot\theta,\tag 2.15$$
where $\cot\theta$ is the boundary parameter appearing in (2.12).
The solution $\varphi(k,x)$ is related to the
vocal-tract radius $r(x)$ as [1,2]
$$\varphi(0,x)=\ds\frac{r(x)}{r(0)},\qquad x\in(0,\ell),\tag 2.16$$
which plays a key role in solving our inverse problem. One can derive
(2.16) by relating (2.10) at $k=0$ and (2.15)
to (2.11) and (2.13). In Section~3 it will become clear that
$\varphi(k,x)$ satisfying (2.10) and (2.15) is the restriction
to $x\in(0,\ell)$ of the regular solution [5,11-13] to
the half-line Schr\"odinger equation
(3.1) with the initial conditions (2.15).

With the help of the
solution $f(k,x)$ and the boundary parameter
$\cot\theta$ appearing in (2.12), let us define
$$F(k):=-i\left[f'(k,0)+(\cot\theta)\,f(k,0)\right],\tag 2.17$$
and hence for the specific boundary parameter in (2.13) we obtain
$$F(k)=-i\left[f'(k,0)-\ds\frac{r'(0)}{r(0)}\,f(k,0)\right],\tag 2.18$$
which will be useful in our analysis of the direct and inverse problems for the
vocal tract.
It is known that [1,2]
$$F(k)=k+O(1),\qquad k\to+\infty,\tag 2.19$$
$$F(-k)=-F(k)^\ast,\qquad k\in\bR.\tag 2.20$$

The pressure and the volume velocity satisfying the two side conditions
mentioned earlier can be evaluated uniquely
and explicitly, as indicated in the following theorem. A proof is omitted
and the reader is referred to [1] for a proof.

\noindent {\bf Theorem 2.1} {\it Assume that the vocal-tract radius
$r(x)$ belongs to class $\Cal A$ specified in Definition~1.1.
Then:}

\item{(a)} {\it The Webster horn equation (2.3) with the boundary
conditions given in (2.7) and
(2.8) has a unique solution, which is given by}
$$P(k,x)=-\displaystyle\frac{c\mu k\,f(-k,x)}
{\pi\,r(0)\,r(x)\,F(-k)},\qquad x\in(0,\ell),\tag 2.21$$
{\it where
$f(k,x)$ is the solution
to (2.10) satisfying (2.14) and
$F(k)$ is the quantity given in (2.18).
Similarly, (2.4) with the boundary conditions given
in (2.6) and (2.9)
is uniquely
solvable, and the solution is given by}
$$V(k,x)=-\displaystyle\frac{i\,r(x)}{r(0)\,F(-k)}
\left[f'(-k,x)-\displaystyle\frac{r'(x)}{r(x)}\,f(-k,x)\right],\qquad
x\in (0,\ell).\tag 2.22$$

\item{(b)} {\it The quantities $P(k,x)$ and $V(k,x)$ are well defined for each
$k\ge 0$ and $x\in(0,\ell).$}

Using (2.14) and (2.20) in (2.21), the absolute pressure at the lips
is seen to be
$$|P(k,\ell)|=\displaystyle\frac{c\mu k}
{\pi\,r(0)\,r(l)\,|F(k)|}, \qquad k\in(0,+\infty).\tag 2.23$$
Having obtained (2.23), we can summarize the steps for a
solution to our direct problem as follows:

\item{(a)} Given $r(x),$ evaluate $q(x)$ defined in (2.11).

\item{(b)} Use $q(x)$ in (2.10), solve the initial-value problem
consisting of the linear homogeneous differential equation (2.10)
with the initial conditions (2.14), and hence recover $f(k,x)$ for $x\in(0,\ell).$

\item{(c)} Using (2.18), evaluate the key quantity $F(k).$

\item{(d)} Determine the absolute pressure $|P(k,\ell)|$ via (2.23).

The procedure described in (a)-(d) above is not necessarily
a straightforward way of solving the direct problem. Its importance,
however, comes from the fact that it relates the vocal-tract radius $r(x)$ to the key quantity $F(k)$
appearing in (2.23), and this is crucial
in the formulation of our inverse problem.

With the goal of providing an alternate solution to our direct problem,
let us define
$$\tilde P(k,x):=\ds\frac{P(k,x)}{P(k,\ell)},\quad \tilde V(k,x):=\ds\frac{V(k,x)}{P(k,\ell)},
\tag 2.24$$
where $P(k,x)$ and $V(k,x)$ are the pressure and volume velocity given in
(2.21) and (2.22), respectively. From (2.6) and the second identity in (2.24) we see that
$$P(k,\ell)=\ds\frac{1}{\tilde V(k,0)}.\tag 2.25$$

\noindent {\bf Theorem 2.2} {\it Assume that the vocal-tract radius
$r(x)$ belongs to class $\Cal A$ specified in Definition~1.1. Then, for each $k>0,$ the pair of quantities
$\tilde P(k,x)$ and $\tilde V(k,x)$ defined in (2.24) forms the unique solution to the initial-value problem
consisting of the first-order system (2.2)
and the initial conditions at $x=\ell$ given by}
$$\tilde P(k,\ell)=1,\quad \tilde V(k,\ell)=\ds\frac{1}{c\mu}\left[A(\ell)+\ds\frac{A'(\ell)}{2ik}\right],
\tag 2.26$$
{\it where $A(x)$ is the area function related to the vocal-tract radius as in (1.1).}

\noindent PROOF: From Theorem~2.1(b) we know that
$P(k,x)$ and $V(k,x)$ are well defined, and hence
 from (2.24) we conclude that $\tilde P(k,x)$ and $\tilde V(k,x)$ are well defined,
 provided $P(k,\ell)\ne 0.$ We confirm later in Theorem~3.2(b) that
 $P(k,\ell)\ne 0$ for $k>0.$
Since the system (2.2) is linear and homogeneous,
the pair $\tilde P(k,x)$ and $\tilde V(k,x)$ defined in (2.24) forms a solution to (2.2)
because we know from Theorem~2.1 that the pair
$P(k,x)$ and $V(k,x)$ forms a solution to (2.2). From the
first identity in (2.24) it is clear that the first equality in (2.26) is satisfied.
Using (2.21) and (2.22) with $x=\ell$ on the right-hand side of (2.24),
with the help of (1.1) and (2.14), we see that the second equality in (2.26) is
also satisfied. \qed

With the help of Theorem~2.2, we can summarize the steps for the alternate
solution to our direct problem as follows:

\item{(a)} Given $r(x),$ with the help of (1.1),
evaluate the initial conditions given in
(2.26).

\item{(b)} Obtain the pair $\tilde P(k,x)$ and $\tilde V(k,x)$ uniquely
by solving the system (2.2) with the initial conditions (2.26).

\item{(c)} Having $\tilde V(k,x)$ at hand, evaluate $\tilde V(k,0).$

\item{(d)} By using (2.25), determine the absolute pressure $|P(k,\ell)|$
as $1/|\tilde V(k,0)|.$

Let us use $P_\infty$ for the asymptotic value of the absolute
pressure when the frequency becomes infinite, i.e.
we let
$$P_\infty:=\displaystyle\lim_{k\to+\infty}\left|
P(k,\ell)\right|.\tag 2.27$$
Using (2.19) and (2.27) in (2.23) we obtain
$$P_\infty=\displaystyle\frac{c\mu}
{\pi\,r(0)\,r(l)},\tag 2.28$$
and hence $P_\infty$ is uniquely determined if we know the product of
the vocal-tract radius at the glottis and at the lips. From (2.23) and (2.28) we see that
$$\ds\frac{|P(k,\ell)|}{P_\infty}=\displaystyle\frac{k}
{|F(k)|}, \qquad k\in(0,+\infty),\tag 2.29$$
which will play an important role in our analysis of the inverse
problem via the Gel'fand-Levitan method.

\vskip 10 pt
\noindent {\bf 3. THE JOST FUNCTION}
\vskip 3 pt

 From (2.27) and (2.29) we know that the knowledge of the absolute pressure at
the lips yields $|F(k)|.$
 In this section we investigate some relevant properties of
 the key function $F(k),$ and those properties are needed in Section~4 in the
 solution of the inverse problem.
 In order to give a physical interpretation to $F(k),$
we extend the Schr\"odinger equation from
$x\in(0,\ell)$ to $x\in(0,+\infty),$ and it turns out that
$F(k)$ is the Jost function associated
with the half-line Schr\"odinger equation and the
boundary condition (2.12)
with the boundary parameter $\cot\theta$ related to the vocal-tract
radius as in (2.13).

The mathematical extension of (2.10) to the half line
gives us the advantage of
relating the acoustic properties pertinent to $x\in(0,\ell)$ to certain quantities related
to the scattering for the half-line Schr\"odinger equation
$$\psi''(k,x)+k^2\psi(k,x)=
q(x)\,\psi(k,x),
\qquad x\in(0,+\infty),\tag 3.1$$
where the potential $q(x)$ is given by (2.11) for $x\in(0,\ell)$
and $q(x)=0$ for $x>\ell.$
In order to have $q(x)=0$ for $x>\ell,$
we choose $r(x)$ as a linear
function of $x$ for $x>\ell.$ Thus, a natural mathematical
extension of $r(x)$ beyond $x=\ell$ is given by
$$r(x)=[r'(\ell)]\, (x-\ell)+r(\ell),\qquad x\ge \ell.\tag 3.2$$

When $r(x)$ belongs to class $\Cal A$ for $x\in(0,\ell),$
the extension described in (3.2) has the advantage that
the corresponding potential $q(x)$ vanishes when $x>\ell$
and it does not contain any singularities or any delta-function
components. Then, $f(k,x)$ appearing in
(2.14) can be extended from the domain
$x\in (0,\ell)$ to $x\in(0,+\infty)$ so that it satisfies
$$f(k,x)=e^{ikx},\quad f'(k,x)=ik\,e^{ikx},\qquad x\ge \ell.\tag 3.3$$
With such an extension, $f(k,x)$ is recognized as being the
Jost solution to the half-line Schr\"odinger equation (3.1) having the
asymptotics $e^{ikx}[1+o(1)]$ as $x\to +\infty.$

Let us impose at $x=0$
the boundary condition given in (2.12)
with $\cot\theta$ as in (2.13).
With the extension from $x\in(0,\ell)$ to
$x\in(0,+\infty),$ the Schr\"odinger equation
(3.1) with the boundary condition
(2.12) yields a selfadjoint differential operator and
the key quantity $F(k)$ given in (2.17)
becomes the corresponding Jost function [5,9,11,12].

The mathematical extension
in (3.2) has a disadvantage as a physical interpretation in the sense that
$r(x)$ becomes negative when $x>[\ell-r(\ell)/r'(\ell)]$
if $r'(\ell)<0.$ Thus, in general one cannot interpret $r(x)$ given in (3.2)
as the physical extension of the
vocal-tract radius to $x\in(0,+\infty).$ One might consider
restricting the physical interpretation of the extension
 from $x\in(0,\ell)$ to only a smaller region in the immediate vicinity
beyond $x=\ell.$

Even though we could extend $r(x)$ from $x\in(0,\ell)$ to
$x\in(\ell,+\infty)$ in many ways other than (3.2),
such other extensions of $r(x)$
may not have a satisfactory physical interpretation beyond $x=\ell$ either,
and the resulting mathematical
theory may be more complicated. In this paper we avoid any issues related to the
modeling of sound propagation outside the vocal tract, and we refer the
reader to [7,13] for further information.

Consider the Schr\"odinger equation (3.1) with the non-Dirichlet boundary condition (2.12),
with $q(x)$ being the real-valued potential given as in
(2.11) but extended from $x\in(0,\ell)$ to $x\in(0,+\infty),$ i.e.
$$q(x)=\ds\frac{r''(x)}{r(x)},\qquad x\in(0,+\infty),\tag 3.4$$
where $r(x)$ is the vocal-tract radius function belonging to class $\Cal A$
for $x\in(0,\ell)$ and
with the extension in (3.2). Recall [5,9,11] that a bound state corresponds to a square-integrable solution to (3.1) and satisfying the boundary condition (2.12).
Let us use $N$ to denote the number of bound states. It is known [5,9,11] that
$N$ is a finite nonnegative integer. For the corresponding Schr\"odinger equation,
let us use
$Z_f$ to denote the number of zeros of $f(0,x)$ in the
interval $[0,+\infty),$ where $f(k,x)$ is the Jost solution
to (3.1) satisfying (3.3). For the same Schr\"odinger operator
let us use $Z_\varphi$ to denote the number of zeros of $\varphi(0,x)$ in the interval
$[0,+\infty),$ where $\varphi(k,x)$ is the regular solution to (3.1)
and satisfying (2.15). From the
first equality in (2.15) we see that $Z_\varphi$ is also equal to the number of zeros of
$\varphi(0,x)$ in the interval $(0,+\infty).$

In the next theorem, we analyze the relationships among $N,$ $Z_f,$ $Z_\varphi,$ and the value of $r'(\ell).$ We use $\bC$ for the complex plane, $\bCp$
for the open upper-half complex plane, and $\bCpb$ for $\bCp\cup\bold R.$

\noindent {\bf Theorem 3.1} {\it Consider the half-line Schr\"odinger equation
(3.1),
with $q(x)$ being the real-valued potential given as in
(3.4), where $r(x)$ is the vocal-tract radius function belonging to class $\Cal A$
for $x\in(0,\ell)$ and
with the extension in (3.2). Let (2.12) be the
boundary condition with the
boundary parameter $\cot\theta$ as in (2.13), $F(k)$ be the
corresponding Jost function appearing in (2.18),
$\varphi(k,x)$ be the regular
solution to (3.1) and satisfying (2.15),
$f(k,x)$ be the Jost solution to (3.1) and satisfying
(3.3). Let $N,$ $Z_\varphi,$ and $Z_f$ denote
the number of bound states, the number of zeros of $\varphi(0,x)$ in the interval
$[0,+\infty),$ and the number of zeros of $f(0,x)$ in the interval
$[0,+\infty),$ respectively. We have the following:}

\item{(a)} {\it The associated Schr\"odinger
operator has no bound states if $r'(\ell)\ge 0$
and it has exactly one bound state if $r'(\ell)<0.$}

\item{(b)} {\it The Jost function $F(k)$ is entire
in $k\in\bC.$ If $r'(\ell)>0,$
then $F(k)$ is nonzero for $k\in\bCpb.$
If $r'(\ell)=0,$
then $F(k)$ is nonzero for $k\in\bCpb\setminus\{0\}$
and it has a simple zero at $k=0.$
If $r'(\ell)<0,$
then $F(k)$ is nonzero for $k\in\bCpb$
with the exception of a single point
on the positive imaginary axis, where that point is a simple zero of $F(k)$
and corresponds to a bound state for the associated Schr\"odinger
operator.}

\item{(c)} {\it We have one of the following four mutually exclusive scenarios:}

\itemitem{(i)} {\it In the first possibility we have}
$$N=0,\quad Z_\varphi=0,\quad Z_f=0,\quad r'(\ell)>0,\quad -i\,F(0)>0,\tag 3.5$$
{\it in which case $\varphi(0,x)$ and $f(0,x)$ are linearly independent on $[0,+\infty).$}

\itemitem{(ii)} {\it In the second possibility we have}
$$N=0,\quad Z_\varphi=0,\quad Z_f=0,\quad r'(\ell)=0,\quad F(0)=0,\tag 3.6$$
{\it in which case $\varphi(0,x)$ and $f(0,x)$ are linearly dependent on $[0,+\infty).$}

\itemitem{(iii)} {\it In the third possibility we have}
$$N=1,\quad Z_\varphi=1,\quad Z_f=0,\quad r'(\ell)<0,\quad -i\,F(0)<0,\tag 3.7$$
{\it in which case $\varphi(0,x)$ and $f(0,x)$ are linearly independent on $[0,+\infty).$}

\itemitem{(iv)} {\it In the fourth possibility we have}
$$N=1,\quad Z_\varphi=1,\quad Z_f=1,\quad r'(\ell)<0,\quad -i\,F(0)<0,\tag 3.8$$
{\it in which case $\varphi(0,x)$ and $f(0,x)$ are linearly independent on $[0,+\infty).$}

\noindent PROOF: Since (c) implies (a), we will prove (a) by proving (c).
Note that the potential $q(x)$ given in (3.4) is real valued,
vanishes when $x>\ell,$ and is integrable as a result of the
facts that $r(x)$
belongs to class $\Cal A$ for $x\in(0,\ell)$
and that the extension of
$r(x)$ to $x\in(\ell,+\infty)$ is given by (3.2).
For such a potential
the corresponding
Jost function $F(k)$ has [5,11,12] the following properties:
$F(k)$ is entire in $k;$ it is nonzero in $k\in\bCpb$ except perhaps for a simple
zero at $k=0$ and a finite number of simple zeros on the
positive imaginary axis in $\bold C,$ with each zero
corresponding to a bound state. We have
$$F(-k)=F(0)-k\,\dot F(0)+O(k^2),\qquad k\to 0 \text{ in } \bC,\tag 3.9$$
where an overdot indicates the $k$-derivative.
Generically we have $F(0)\ne 0,$ and this happens when
$\varphi(0,x)$ is unbounded on $x\in[0,+\infty).$
In the exceptional case we have $F(0)=0,$
and this happens
when $\varphi(0,x)$ is bounded on $x\in[0,+\infty).$
Thus, the proof of (b) will be complete if we prove (c).
Let us now turn to the proof of (c).
Note that (2.16) has the extension
$$\varphi(0,x)=\ds\frac{r(x)}{r(0)},\qquad x\in(0,+\infty),\tag 3.10$$
where $r(x)$ for $x>\ell$ is given by (3.2).
Using Definition~1.1 and (3.2) in (3.10) we conclude that $Z_\varphi=0$ if
$r'(\ell)\ge 0$ and that $Z_\varphi=1$ if $r'(\ell)<0.$ Furthermore, when $Z_\varphi=1,$
the zero of $\varphi(0,x)$ must occur in $(\ell,+\infty).$ From (3.3)
we see that
$$f(0,x)=1,\quad f'(0,x)=0,\qquad x\in[\ell,+\infty),\tag 3.11$$
and hence any possible zeros of $f(0,x)$
can only occur in $[0,\ell).$ We must have either $Z_f=0$ or $Z_f=1,$ because if $f(0,x)$ had two or more zeros
in $[0,\ell)$ then there would have to be at least one zero of $\varphi(0,x)$ in
$(0,\ell)$ as a result of the interlacing property [10,20] of the zeros of
$\varphi(0,x)$ and $f(0,x).$ It is already known [10,20] that
there are no further possibilities other than
the two possibilities $N=Z_f$ and
$N=Z_f+1.$ Thus, $N$ cannot exceed $2.$ We will now
prove that we cannot have $N=2$ and hence we must have either $N=1$ or $N=0.$
In terms of the Jost function $F(k),$ let us define
$$H(\beta):=-i\,F(i\beta).\tag 3.12$$
 From (2.19) it follows that $H(\beta)=\beta+O(1)$ as $\beta \to +\infty,$ and
it is known [5,9,11] that each bound-state zero of
$F(k)$ corresponds to a simple zero of $H(\beta)$ in the interval
$\beta\in(0,+\infty).$
It is known [10,20] that
there are no further possibilities other than
the two possibilities $N=Z_\varphi$ and $N=Z_\varphi+1.$ Hence, if we had
$N=2$ then we would have to have
$Z_\varphi=1$ because we already know that we cannot have $Z_\varphi=2.$
 From (2.15), (2.17), and (3.12) we get
$$H(0)=f(0,x)\,\varphi'(0,x)-f'(0,x)\,\varphi(0,x).\tag 3.13$$
We observe that the right-hand side in (3.13) is the Wronskian of two solutions to
(3.1) at $k=0.$ That Wronskian is known [5,9,11] to be
independent of $x$ and can be evaluated at
any $x$-value. Using (3.2), (3.10), and (3.11) in (3.13) we obtain
$$H(0)=\ds\frac{r'(\ell)}{r(0)}.\tag 3.14$$
As we have seen, when $N=2$ the
only possibility is $Z_\varphi=1,$ and hence
we must have $r'(\ell)<0,$ yielding $H(0)<0$
in (3.14). On the other hand, with $N=2$ the graph of $H(\beta)$ would have two simple zeros in $\beta\in(0,+\infty)$ and hence we would have $H(0)\ge 0.$ This contradiction shows that
we cannot have $N=2.$ Thus, we must have either $N=1$ or $N=0.$
When $N=0,$ since neither $Z_f$ nor $Z_\varphi$ can exceed
$N,$ we either have (3.5) with $H(0)>0$ or we have (3.6) with $H(0)=0,$
where by (3.12) we know that
$H(0)>0$ is equivalent to $-i\, F(0)>0$ and that
$H(0)=0$ is equivalent to $F(0)=0.$ When $N=1,$
we can either have the possibility in (3.7) or the possibility in (3.8). In other words,
we cannot have the possibility
$$N=1,\quad Z_\varphi=0,\quad Z_f=1.\tag 3.15$$
If we had (3.15), then we would have $r'(\ell)\ge 0$ due to $Z_\varphi=0,$ but
we would also have $H(0)\le 0$ due to $N=1.$ Thus, if we had (3.15), then
we would have to have $H(0)=0,$ which, by (3.13), would imply that $f(0,x)$ and $\varphi(0,x)$ would
have to be linearly dependent on $[0,+\infty).$ However, that linear dependence would
require $Z_\varphi=Z_f,$ contradicting (3.15). Let us remark that we cannot have
$H(0)=0$ in the third possibility in (c) because we already have $Z_\varphi\ne Z_f$ there. Furthermore,
we cannot have $H(0)=0$ in the fourth possibility
in (c) because the zero of $f(0,x)$ must occur in
$[0,\ell)$ and the zero of $\varphi(0,x)$ must occur in
$(\ell,+\infty).$ Thus, the proof of (c) is complete. \qed

Two of the authors recently analyzed the inverse problem
of recovery of a compactly-supported potential on the half line and of the boundary condition
when the available input data set consists [4] of
the absolute value of the Jost function, without having any
explicit information on the bound states.
The analysis
in [4] was actually motivated by the inverse scattering problem
of recovery of the vocal-tract radius from the absolute
pressure at the lips. The results given in [4] in the special case
of one bound state are directly relevant to the study in our current paper.

In the following theorem we provide some relevant properties of the
pressure and volume velocity in the vocal tract.

\noindent {\bf Theorem 3.2} {\it Assume that $r(x)$ belongs to class $\Cal A$ for
$x\in(0,\ell)$ and
has the extension given in (3.2).
Let $P(k,x)$ and $V(k,x)$ be the corresponding pressure
and the volume velocity, given in (2.21) and (2.22), respectively. Then:}

\item{(a)} {\it For each fixed $k\ge 0,$ the quantities
$P(k,x)$ and $V(k,x)$ are continuous in $x\in(0,\ell).$}

\item{(b)} {\it The quantity $P(k,\ell)$ is nonzero for $k>0,$ and
$P(k,\ell)$ is either nonzero at $k=0$ or it has a simple zero at $k=0.$}

\noindent PROOF: From (2.16), (3.2), (3.4), and the
properties of $r(x)$ listed in Definition~1.1, it follows that the potential
$q(x)$ defined in (3.4) is integrable and vanishes when $x>\ell.$
Consequently [5,11,12], for each fixed $x\in(0,\ell)$ the corresponding
Jost solution $f(k,x)$ appearing in (2.21) and
$f'(k,x)$ are entire in $k$
and for each $k\in\bC$ the quantities $f(k,x)$ and $f'(k,x)$
 are continuous in $x\in(0,\ell).$
 From Theorem~3.1 we know that $1/F(k)$ and $k/F(k)$ are nonzero
 for $k\in\bR\setminus\{0\}.$ From Definition~1.1 we
 have the continuity of $r(x)$ and of $r'(x)$ and the positivity of
 $r(x)$ for $x\in(0,\ell).$ Thus, from (2.21) and (2.22) we conclude
the continuity of $P(k,x)$ and $V(k,x)$ in $x\in(0,\ell)$
for each $k>0.$ Let us now prove the continuity
of $P(0,x)$ and $V(0,x)$ in $x\in(0,\ell).$
By letting $k=0$ in (2.2) we see that $P'(0,x)=0$ and
$V'(0,x)=0$ for $x\in(0,\ell).$ Thus,
$P(0,x)$ and $V(0,x)$ are both constants
and independent of $x.$
Hence, their values can be evaluated at
$x=0$ or at $x=\ell.$ As a result, with the help of
(2.6) we conclude that
$$V(0,x)\equiv 1,\tag 3.16$$
which confirms the continuity of $V(0,x).$
Because $F(k),$ $f(k,x),$ and $f'(k,x)$ are analytic
in $k,$ from (2.21) and (2.22) we conclude that
$P(k,x)$ and $V(k,x)$ are analytic in $k$ for each
fixed $x\in(0,\ell).$ Thus, $P(0,x)$ and $V(0,x)$ can also
be obtained
by letting
$k\to 0$ in (2.21) and (2.22), respectively.
With the help of (3.3) and (3.9), from (2.22) we obtain
$$V(k,\ell)=\ds\frac{-i\,r(\ell)\left[-ik-r'(\ell)/r(\ell)\right]\left[1-ik\ell+O(k^2)\right]}
{r(0)\left[F(0)-k\,\dot F(0)+O(k^2)\right]},
\qquad k\to 0 \text{ in } \bC.\tag 3.17$$
By Theorem~3.1(b) we know that $F(0)=0$ if and only if $r'(\ell)=0.$
Hence, from (3.17) we conclude that
$$V(0,\ell)=\cases \ds\frac{i\,r'(\ell)}{r(0)\,F(0)},\qquad \text{ if } F(0)\ne 0,\\
\stretch
\ds\frac{r(\ell)}{r(0)\,\dot F(0)},\qquad \text{ if } F(0)= 0.\endcases\tag 3.18$$
Comparing (3.16) and (3.18) we conclude that
$$F(0)=\ds\frac{i\,r'(\ell)}{r(0)},\tag 3.19$$
$$\dot F(0)=\ds\frac{r(\ell)}{r(0)},\qquad \text{ if } F(0)=0.\tag 3.20$$
Let us remark that (3.19) is equivalent to (3.14), which is seen with the
help of (3.12).
We
now turn to the analysis of $P(k,x)$ as $k\to 0.$
Using (3.9) and the analyticity of $f(k,x)$
 in $k$ at $k=0,$ from (2.21) we obtain
$$P(k,x)=-\ds\frac{c\mu k\left[f(0,x)-k\,\dot f(0,x)+O(k^2)\right]}
{\pi\,r(0)\,r(x)\left[ F(0)-k\,\dot F(0)+O(k^2)\right]},\qquad
k\to 0{\text{ in }}\bC.\tag 3.21$$
If $F(0)\ne 0,$ then from (3.21) we get
$$P(k,x)=-\ds\frac{c\mu\,f(0,x)}{\pi \,r(0)\,r(x)\, F(0)}\,k+O(k^2)
 ,\qquad
k\to 0{\text{ in }}\bC,\tag 3.22$$
yielding
$$P(0,x)\equiv 0,\qquad \text{ if } F(0)\ne 0.\tag 3.23$$
On the other hand, if $F(0)=0$ then
$\dot F(0)\ne 0$ because of the simplicity of the zero of $F(k)$ at $k=0,$ as stated
in Theorem~3.1(b). Thus, if $F(0)=0,$ then from (3.21) we obtain
$$P(k,x)=\ds\frac{c\mu\,f(0,x)}{\pi \,r(0)\,r(x)\, \dot
F(0)}+O(k)
 ,\qquad
k\to 0{\text{ in }} \bC.\tag 3.24$$
 From (3.24) we conclude that
$$P(0,x)= \ds\frac{c\mu\,f(0,x)}{\pi \,r(0)\,r(x)\, \dot
F(0)},\qquad \text{ if } F(0)=0.\tag 3.25$$
We already know that $P(0,x)$ must be independent
of $x,$ and hence we can evaluate the right-hand side of (3.25)
at $x=\ell$ with the help of (3.3) and (3.20).
We then obtain
$$P(0,x)\equiv \ds\frac{c\mu}{\pi\,r(\ell)^2},\qquad \text{ if } F(0)=0.\tag 3.26$$
 Therefore, from (3.16), (3.23), and (3.26) we conclude the
 continuity of $P(k,x)$ in $x\in(0,\ell)$ also when $k=0.$
Thus, the proof of (a) is complete. Let us now turn to the proof
of (b). Using the first
 equality in (2.14) in (2.21) we get
 $$P(k,\ell)=-\ds\frac{c\mu k\, e^{-ik\ell}}{\pi\,r(0)\,r(\ell)\,F(-k)}.\tag 3.27$$
 By Theorem~3.1(b), the quantity $k/F(k)$ is nonzero for
 $k\in\bR\setminus\{0\}.$ From Definition~1.1 we have $r(0)\,r(\ell)>0.$
 Hence, from (3.27) we conclude that $P(k,\ell)$ does not vanish
 when $k>0.$ From (3.22), (3.23), and (3.25), we conclude that
 $P(k,\ell)$ vanishes linearly in $k$ as $k\to 0$ if
 $F(0)\ne 0$ and that
 $P(0,\ell)\ne 0$ if $F(0)=0.$ Thus, the proof of (b) is complete. \qed

 Let us remark that it is possible to provide an alternate proof
 that
 the right-hand side of (3.25) is independent of $x$
 by proceeding as follows. It is known [5,11,12] that
$$\varphi(k,x)=\ds\frac{1}{2k}\left[F(k)\,f(-k,x)-F(-k)\,f(k,x)\right],
\qquad x\in(0,+\infty),\tag 3.28$$
 from which, by letting $k\to 0,$ we obtain
$$\varphi(0,x)=\dot F(0)\, f(0,x)-F(0)\,\dot f(0,x).\tag 3.29$$
If $F(0),$ then (3.29) reduces to
$$\varphi(0,x)=\dot F(0)\, f(0,x),\qquad \text{ if } F(0)=0.\tag 3.30$$
Using (2.16) on the left-hand side in (3.30) we conclude that
for $x\in(0,\ell)$ we get
$$\ds\frac{r(x)}{r(0)}=\dot F(0)\, f(0,x),\qquad \text{ if } F(0)=0.\tag 3.31$$
Finally, using (3.31) in (3.25), we obtain
$$P(0,x)\equiv \ds\frac{c\mu}{\pi\,r(0)^2 \dot F(0)^2},\qquad \text{ if } F(0)=0,$$
which is seen equivalent to (3.26) with the help of (3.20).

In Theorem~3.1 we have seen that the
sign of $r'(\ell)$ plays a crucial role. The next proposition shows that the
sign of $r'(\ell)$ is actually
related to the small-$k$ limit of $P(k,\ell),$ the pressure at the lips.

\noindent {\bf Proposition 3.3} {\it Assume
that the vocal-tract radius
$r(x)$ belongs
 to class $\Cal A$ for $x\in(0,\ell)$ and has the extension given in (3.2).
 Let $P(k,\ell)$ be the corresponding
 pressure at the lips given in (3.27). Consider the corresponding Schr\"odinger operator
 where the potential $q(x)$ is related to $r(x)$ as in
(3.4) and the boundary parameter $\cot\theta$
appearing in (2.12) is related to
$r(x)$ as in (2.13), and let $F(k)$ be the corresponding
Jost function given in (2.18). Then:}

 \item{(a)} {\it We have $r'(\ell)=0$ if and only if $P(0,\ell)\ne 0.$}

 \item{(b)} {\it We have $r'(\ell)>0$ if and only if $P(0,\ell)= 0$
 and $i \dot P(0,\ell)<0.$}

\item{(c)} {\it We have $r'(\ell)<0$ if and only if $P(0,\ell)= 0$
 and $i \dot P(0,\ell)>0.$}

 \noindent PROOF: It is enough to prove the if-parts in (a)-(c) because the three
 outcomes in (a)-(c) are mutually exclusive and cover all possibilities.
  Using (2.28) in (3.27) we get
$$P(k,\ell)=\ds\frac{-P_\infty\, k\,e^{-ik\ell}}
{F(-k)},\tag 3.32$$
where $P_\infty$ is the positive constant defined in (2.27).
 By (3.6) in Theorem~3.1(b) we know that $r'(\ell)=0$ implies that $F(0)=0,$ and hence
using (3.9) in
(3.32) we get
$$P(k,\ell)=\ds\frac{P_\infty}{\dot F(0)}+O(k),
\qquad k\to 0 \text{ in } \bC,\tag 3.33$$
which implies that $P(0,\ell)$ is nonzero and equal to $P_\infty/\dot F(0),$
proving the if-part in (a).
By (3.5) in Theorem~3.1(b), when $r'(\ell)>0$ we have $F(0)\ne 0$
and $-i\,F(0)>0.$ In that case, using (3.9) in (3.32) we get
$$P(k,\ell)=\ds\frac{-P_\infty\, k}{F(0)}+O(k^2),
\qquad k\to 0 \text{ in } \bC,\tag 3.34$$
which implies that $P(0,\ell)=0$ and $\dot P(0,\ell)=-P_\infty/F(0).$
Thus, the sign of $i \dot P(0,\ell)$ is the same as the sign of
$i F(0),$ which is negative. Hence, the if-part in (b) is proved.
Finally, by (3.7) and (3.8) in Theorem~3.1(b), when $r'(\ell)<0$ we have $F(0)\ne 0$
and $-i\,F(0)<0.$ Thus, (3.34) implies that $P(0,\ell)=0$ and $\dot P(0,\ell)=-P_\infty/F(0).$
Therefore, the sign of $i \dot P(0,\ell)$ is the same as the sign of
$i F(0),$ which is positive. Hence, the if-part in (c) is proved. \qed

In the next theorem we present the large-frequency behavior of the absolute pressure
under some further restriction on the vocal-tract radius function.

\noindent {\bf Theorem 3.4} {\it Consider the Schr\"odinger equation
on the half line $x\in(0,+\infty)$ given in (3.1) with the boundary
condition in (2.12), and assume that
the potential $q(x)$ is real valued, vanishes when
$x>\ell,$ and is continuous in $x\in(0,\ell)$
with finite limits $q(0^+)$ and $q(\ell^-).$
Let $F(k)$ be the corresponding Jost function defined in (2.17).
Then, we have}
$$\ds\frac{k^2}{|F(k)|^2}=1-\ds\frac{1}{k^2}\left[\cot^2\theta-\ds\frac{q(0^+)}{2}+
\ds\frac{q(\ell^-)}{2}\,\cos(2k\ell)\right]+O\left(\ds\frac{1}{k^3}\right),
\qquad k\to\pm\infty.\tag 3.35$$
{\it Consequently, if the vocal-tract radius $r(x)$ belongs to class
$\Cal A$ and we further assume that
$r''(x)$ is continuous for $x\in(0,\ell)$ with finite limits
$r''(0^+)$ and $r''(\ell^-),$ then the absolute pressure $|P(k,\ell)|$
at the lips
has the large-frequency behavior}
$$\ds\frac{|P(k,\ell)|^2}{P_\infty^2}=1-\ds\frac{1}{k^2}\left[\ds\frac{r'(0)^2}{r(0)^2}
-\ds\frac{r''(0^+)}{2\,r(0)}+
\ds\frac{r''(\ell^-)}{2\,r(\ell)}\,\cos(2k\ell)\right]+O\left(\ds\frac{1}{k^3}\right),
\qquad k\to\pm\infty,\tag 3.36$$
{\it where $P_\infty$ is the positive constant given in (2.28).}

\noindent PROOF: Let
$$m(k,x):=e^{-ikx}f(k,x),\tag 3.37$$
 where $f(k,x)$ is the Jost solution to (3.1)
appearing in (3.3). Under the stated conditions on the potential $q(x),$ we have
the large-$k$ estimates given in (7.5) of [3] and in the first equation in (7.7) of [3],
namely
$$m(k,0)=1-\ds\frac{a_1}{2ik}-\ds\frac{a_1^2-a_2}{8k^2}+O\left(\ds\frac{1}{k^3}\right),
\qquad k\to\pm\infty,\tag 3.38$$
$$m'(k,0)=\ds\frac{a_2}{2ik}+O\left(\ds\frac{1}{k^2}\right),
\qquad k\to\pm\infty,\tag 3.39$$
where we have defined
$$a_1:=\int_0^\ell dx\,q(x),\qquad a_2:=q(0^+)-q(\ell^-)\,e^{2ik\ell}.\tag 3.40$$
Note that we can express (2.17) in terms of $m(k,0)$ and $m'(k,0)$ as
$$F(k)=(k-i\,\cot\theta)\,m(k,0)-i\,m'(k,0).\tag 3.41$$
Using (3.38)-(3.40) in (3.41) we obtain
$$F(k)=k+i\left[\ds\frac{a_1}{2}-\cot\theta\right]+\ds\frac{1}{k}
\left[-\ds\frac{a_1^2}{8}-\ds\frac{a_2}{4}+\ds\frac{a_1}{2}\,\cot\theta\right]+
O\left(\ds\frac{1}{k^2}\right),
\qquad k\to\pm\infty.\tag 3.42$$
With the help of (2.20), after some simplifications, from (3.42) we obtain
(3.35). Then, from (3.35), with the help of (2.13), (2.29), and (3.4), we get
(3.36). \qed

\vskip 10 pt
\noindent {\bf 4. THE SOLUTION TO THE INVERSE PROBLEM}
\vskip 3 pt

In this section we consider the inverse problem of
recovery of the vocal-tract radius $r(x)$ for $x\in(0,\ell)$
 from the absolute pressure
at the lips,
i.e. from $|P(k,\ell)|$ known for $k>0.$
As a result of (2.29), we relate our
inverse problem to the recovery of the
potential and of the boundary parameter for the
half-line Schr\"odinger equation from the absolute value
of the corresponding Jost function, where there is at most one
bound state. We show that there are exactly $M+1$ candidates for the
vocal-tract radius function for a given input data set
consisting of the absolute pressure at the lips, where
$M$ is a nonnegative integer uniquely determined by
our input data set. The value of $M$ is equal to the maximal
number of eligible resonances [4] associated with
the half-line Schr\"odinger equation (3.1) with
the boundary condition (2.12) with
$\cot\theta$ as in (2.13).
One of the $M+1$ candidates corresponds to a potential
with no bound states and to a vocal-tract radius with $r'(\ell)\ge 0.$
Each of the remaining $M$ candidates corresponds to a potential with
one bound state, as there are $M$ distinct choices for a bound state.
Each of these $M$ choices is also a candidate for a vocal-tract radius
with $r'(\ell)<0$ and each $r(x)$ having
exactly one zero in the interval $x\in (0,+\infty).$ Since we require that
the corresponding vocal-tract radius must be positive for $x\in[0,\ell],$
we only allow those candidates for the vocal tract where the
extension of the radius becomes zero
in the interval $x\in(\ell,+\infty)$ and we label the remaining ones as
inadmissible. We provide an equivalent admissibility criterion for each of the $M$ candidates.

We present two recovery methods to obtain each of the $M+1$
candidates for the vocal-tract radius from the absolute pressure at the
lips. The first method is based
on the Gel'fand-Levitan method [5,9,11,12] and the second is based on
the Marchenko method [5,11,12].
We indicate that a Darboux transformation [4] can also be used to obtain
the remaining $M$ candidates for
radius functions after recovering one of the
candidates via the Gel'fand-Levitan method or
the Marchenko method with our input data set.


We elaborate on the number $M$ in the proof of Theorem~4.1.
Let us remark that, in theory,
 $M$ can be infinite but under some further minimal assumption
  on the potential, it is guaranteed that $M$ is
 finite. For example, from Proposition 7 of [21] it follows that, if $q(x)\ge 0,$
 or $q(x)\le 0,$ in some neighborhood of $x=\ell,$ then $M$ is finite.
Recall that we assume that the vocal-tract radius satisfies the conditions
stated in Definition~1.1 and (3.2).
 Consequently, with the help of (3.4), we see that
 the finiteness of $M$ is guaranteed
 if we further assume that $r''(x)$ is continuous in $x\in(\ell-\epsilon,\ell)$
 for some positive $\epsilon$ and that either $r''(x)\ge 0$ or
 $r''(x)\le 0$ for $x\in(\ell-\epsilon,\ell).$

The following theorem deals with eligible resonances corresponding to
a compactly-supported potential with one bound state.

\noindent {\bf Theorem 4.1} {\it Consider the Schr\"odinger equation
on the half line $x\in(0,+\infty)$ with a real-valued
potential $q(x),$ where $q(x)$ is integrable on $x\in(0,\ell)$
and vanishes when $x>\ell.$ Supplement the Schr\"odinger equation
with the boundary condition
given by (2.12), and let
$F(k)$ given in (2.17) be the corresponding Jost function
for the associated Schr\"odinger operator.
Further, suppose that there is exactly one bound state for that Schr\"odinger operator.
Let our input data set consist of $|F(k)|$ for $k>0.$ Then:}

\item{(a)} {\it The maximal number of eligible resonances, $M,$ is uniquely determined
by our input data set. The value of $M$ is at least $1.$}

\item{(b)} {\it  The $k$-values corresponding to the eligible resonances,
denoted by the ordered set $\{-i\beta_1,\dots,-i\beta_M\},$ are
uniquely determined
by our input data set.}

\item{(c)} {\it Corresponding to our input data set,
there are exactly $M$ sets $\{q_j(x),F_j(k)\}$ for
$j=1,\dots,M,$ each consisting of a compactly-supported
potential $q_j(x)$ and the Jost function
$F_j(k)$ having exactly one bound-state zero.
Each of these $M$ sets is uniquely determined by
our input data set.}

\item{(d)} {\it Our input data set corresponds to
exactly $M$ sets $\{\beta_j,\cot\theta_j,
\varphi_j(k,x),f_j(k,x),g_j\}$
for
$j=1,\dots,M,$ where $k=i\beta_j$ is the bound-state
wavenumber, $\cot\theta_j$ is the boundary parameter
appearing in (2.12), $\varphi_j(k,x)$ is
the regular solution satisfying (2.15) with
the boundary parameter $\cot\theta_j,$
$f_j(k,x)$ is the Jost solution satisfying (2.14),
and $g_j$ is the Gel'fand-Levitan
norming constant defined as}
$$g_j:=\ds\frac{1}{\ds\sqrt{\ds\int_0^\infty dx\,
[\varphi_j(i\beta_j,x)]^2}}.\tag 4.1$$
{\it The collection of these $M$ sets is uniquely determined by
our input data set.}

\noindent PROOF: Because of (2.20) our input data
set is equivalent to
having $|F(k)|$ for $k\in\bR.$ Let
$$\overset{\circ}\to F(k):=k\,\exp\left(\ds\frac{-1}{\pi i}
\ds\int_{-\infty}^\infty dt\, \ds\frac{\log|t/F(t)|}{
t-k-i0^+}\right),\tag 4.2$$
where $i0^+$ indicates that the value for $k\in\bR$ must be obtained
as a limit from $\bCp.$ It is known [5] that $\overset{\circ}\to F(k)$
corresponds to the Jost function of the half-line Schr\"odinger
operator with the boundary condition given in (2.12) for some boundary
parameter $\cot\overset{\circ}\to \theta$ and for a potential
$\overset{\circ}\to q(x)$ without any bound states in such a way that $\overset{\circ}\to q(x)$
vanishes [4] when $x>\ell.$ Let $\overset{\circ}\to \varphi(k,x)$ be the regular solution
corresponding to $\overset{\circ}\to F(k).$ It is known [5,9,11,12] that
$\overset{\circ}\to \varphi(k,x)$ is uniquely determined by our input data set
and it satisfies
$$\overset{\circ}\to \varphi''(k,x)+k^2\,\overset{\circ}\to \varphi(k,x)=\overset{\circ}\to q(x)\,\overset{\circ}\to \varphi(k,x),
\qquad x\in(0,+\infty),$$
$$\overset{\circ}\to \varphi(k,0)=1,\quad \overset{\circ}\to \varphi'(k,0)=
-\cot\overset{\circ}\to \theta,\tag 4.3$$
for a uniquely determined value of $\cot\overset{\circ}\to \theta.$
In fact, the construction of $\overset{\circ}\to q(x)$ and
$\overset{\circ}\to \varphi(k,x)$ from our
input data set can be accomplished by the Gel'fand-Levitan procedure [5,9,11,12] as
follows. First, we form the Gel'fand-Levitan kernel given by
$$\overset{\circ}\to G(x,y):=\ds\frac{2}{\pi}\ds\int_0^\infty dk\,
\left[\ds\frac{k^2}{|F(k)|^2}-1\right]
\left(\cos kx\right)\left(\cos ky\right).\tag 4.4$$
Next, we use $\overset{\circ}\to G(x,y)$ as input to the Gel'fand-Levitan equation
$$\overset{\circ}\to h(x,y)+\overset{\circ}\to G(x,y)
+\int_0^x dz\,\overset{\circ}\to h(x,z)\, \overset{\circ}\to G(z,y)=0,
\qquad 0\le y<x.\tag 4.5$$
The potential $\overset{\circ}\to q(x),$ the boundary parameter
$\cot\overset{\circ}\to \theta,$ and the regular solution
$\overset{\circ}\to \varphi(k,x)$ are obtained via
$$\overset{\circ}\to q(x)=2\ds \frac{d\overset{\circ}\to h(x,x)}{dx},
\quad \cot\overset{\circ}\to \theta=-\overset{\circ}\to h(0,0),\quad
\overset{\circ}\to \varphi(k,x)=\cos kx+\ds\int_0^x dy\,
\overset{\circ}\to h(x,y)\,\cos ky.\tag 4.6$$
It is known [5,9,11,12] that $\overset{\circ}\to F(k)$ is entire
because $q(x)$ is assumed to have a compact support. The resonances
correspond to the zeros of $\overset{\circ}\to F(k)$ in the lower-half
complex plane. The imaginary resonances correspond to the
zeros of $\overset{\circ}\to F(k)$ on the negative imaginary axis in
$\bC.$ From (3.52)
of [4] it follows that
the eligible resonances are those imaginary resonances at which
$\overset{\circ}\to F(k)$ vanishes and $d\overset{\circ}\to
F(k)/dk$ has a positive value. In other words, $k=-i\beta_j$
for some $\beta_j>0$ corresponds to an eligible resonance if and only if
$$\overset{\circ}\to F(-i\beta_j)=0,\quad
\ds\frac{d \overset{\circ}\to F(-i\beta_j)}{dk}>0.\tag 4.7$$
Because of the assumption that $q(x)$ has one bound state,
we already know that the number of positive $\beta_j$-values
satisfying (4.7) is at least one. Let $M$ denote
the total number of such $\beta_j$-values satisfying (4.7). In [4] the
number $M$ is called the maximal number of eligible resonances.
Because $\overset{\circ}\to F(k)$ is entire, the value of
$M$ and the set $\{\beta_j\}_{j=1}^M$ are uniquely determined
by $\overset{\circ}\to F(k).$ Since our input data set
uniquely determines $\overset{\circ}\to F(k),$ it follows that
our input data set uniquely determines $M$ and $\{-i\beta_1,\dots,
-i\beta_M\}.$ Thus, we have proved (a) and (b). Let us now prove
(c). Suppose we would like to add a bound state to
$\overset{\circ}\to q(x)$ in such a way that the resulting
potential is compactly supported. It is known [4]
that such a bound state must occur at $k=i\beta_j$ for one
of the $j$-values with
$j=1,\dots,M,$ and hence there are exactly $M$ ways to choose
the set $\{q_j(x),F_j(k)\}.$ For each choice of $\beta_j,$
the corresponding Jost function
$F_j(k)$ is uniquely determined because it is related to
$\overset{\circ}\to F(k)$ as
$$F_j(k)=\ds\frac{k-i\beta_j}{k+i\beta_j}\,\overset{\circ}\to F(k).
\tag 4.8$$
Since the $\beta_j$-values are real, from (4.8) it follows that
$$|F_j(k)|=|\overset{\circ}\to F(k)|,\qquad k\in\bR.\tag 4.9$$
The potential $q_j(x)$ is uniquely determined with the help of (3.2)
of [4] as
$$q_j(x)=\overset{\circ}\to q(x)-\ds\frac{d}{dx}\left[
\ds\frac{2g^2_j\, \left[\overset{\circ}\to \varphi(i\beta_j,x)\right]^2}
{1+g^2_j\ds\int_0^x dy\,\left[\overset{\circ}\to \varphi(i\beta_j,y)\right]^2}
\right],$$
where the positive constant $g_j,$ known as the Gel'fand-Levitan norming constant
associated with the bound state $k=i\beta_j,$
is obtained with the help of (3.19) of [4] via
$$g_j^2=\ds\frac{2\beta_j}{\overset{\circ}\to \varphi(i\beta_j,\ell)^2-2\beta_j\ds\int_0^\ell dy\,
\overset{\circ}\to \varphi(i\beta_j,y)^2}.\tag 4.10$$
Thus, (c) is proved. Let us now prove (d). We already know from (c)
that our input data set corresponds to
$M$ sets $\{q_j(x),F_j(k)\}$
for $j=1,\dots,M,$ each of which
is associated with a specific choice of $\beta_j.$
 From (4.10) we know that
the Gel'fand-Levitan norming constant $g_j$ at the bound state
$k=i\beta_j$ corresponding to the Jost function $F_j(k)$ is also uniquely
determined by our input data set.
Furthermore, from (4.9) we know that $|F_j(k)|$ for $k\in\bR$
is uniquely determined by our input data set.
By the Gel'fand-Levitan procedure, we can uniquely determine the potential
$q_j(x)$ and the boundary parameter $\cot\theta_j$ by proceeding
as in (4.5) and (4.6).
We first construct the Gel'fand-Levitan kernel given by
$$G_j(x,y):=\ds\frac{2}{\pi}\ds\int_0^\infty dk\,\left[\ds\frac{k^2}{|F(k)|^2}-1\right]
\left(\cos kx\right)\left(\cos ky\right)+g_j^2 \left(\cosh \beta_j x\right)
\left(\cosh \beta_j y\right).\tag 4.11$$
Using $G_j(x,y)$ as input to the Gel'fand-Levitan equation
$$h_j(x,y)+G_j(x,y)+\int_0^x dz\,h_j(x,z)\, G_j(z,y)=0,
\qquad 0\le y<x,\tag 4.12$$
we obtain $h_j(x,y),$ from which
the potential $q_j(x),$ the boundary parameter $\cot\theta_j,$
and the regular solution $\varphi_j(k,x)$ are constructed as
$$q_j(x)=2\ds \frac{d h_j(x,x)}{dx},\quad \cot\theta_j=-h_j(0,0),\quad
\varphi_j(k,x)=\cos kx+\ds\int_0^x dy\,
h_j(x,y)\,\cos ky,\tag 4.13$$
where $\varphi_j(k,x)$ is the regular solution to the
Schr\"odinger equation given by
$$\varphi_j''(k,x)+k^2\,\varphi_j(k,x)=q_j(x)\,\varphi_j(k,x),
\qquad x\in(0,+\infty),\tag 4.14$$
and satisfying the initial conditions
$$\varphi_j(k,0)=1,\quad \varphi_j'(k,0)=
-\cot\theta_j.\tag 4.15$$
The Jost solution $f_j(k,x)$ is uniquely determined by solving the
Schr\"odinger equation (4.14) with the initial conditions given in (3.3).
Thus, we have proved (d). \qed

Along with the Gel'fand-Levitan norming constant
$g_j$ given in (4.1)
we have the Marchenko norming constant defined as [5,11,12]
$$m_j:=\ds\frac{1}{\ds\sqrt{\ds\int_0^\infty dx\,
[f_j(i\beta_j,x)]^2}},\tag 4.16$$
where $f_j(k,x)$ is the Jost solution appearing
in Theorem~4.1(d). Let us remark that using the results in [4]
the Gel'fand-Levitan norming constant
$g_j$ appearing in (4.1) and the Marchenko norming
constant $m_j$ appearing in (4.16) can be expressed
explicitly in terms of $\overset{\circ}\to F(k)$ or
the Jost function $F_j(k)$
appearing in (4.8) and the value of $\beta_j.$
The results are given in the following theorem.

\noindent {\bf Theorem 4.2} {\it Consider the Schr\"odinger equation
(4.14) with a real-valued
potential $q_j(x),$ where $q_j(x)$ is integrable on $x\in(0,\ell)$
and vanishes when $x>\ell.$ With the boundary condition
given by (2.12) but $\cot\theta$ replaced by
$\cot\theta_j$ there, let
$F_j(k)$ given as in (2.17) be the corresponding Jost function
for the associated Schr\"odinger operator.
Further, suppose that there is exactly one bound state for that Schr\"odinger operator
and that the bound state occurs at $k=i\beta_j.$
Let $\overset{\circ}\to F(k)$ be the quantity appearing in (4.8).
Then, the Marchenko norming constant
$m_j$ defined in (4.16) is related to $F_j(k)$ and $\overset{\circ}\to F(k)$ as}
$$m_j^2=\ds\frac{i\, F_j(-i\beta_j)}{\ds\frac{dF_j(k)}{dk}\bigg|_{k=i\beta_j}
}=\ds\frac{4i\beta_j^2}{\overset{\circ}\to F_j(i\beta_j)}\,
\ds\frac{d \overset{\circ}\to F_j(k)}{dk}\bigg|_{k=-i\beta_j}
,\tag 4.17$$
{\it and similarly, the Gel'fand-Levitan norming constant
$g_j$ defined in (4.1) is related to $F_j(k)$ and $\overset{\circ}\to F(k)$ as}
$$g_j^2=\ds\frac{-4i\beta_j^2}{F_j(-i\beta_j)\,\ds\frac{dF_j(k)}{dk}\bigg|_{k=i\beta_j}
}=\ds\frac{4i\beta_j^2}{\overset{\circ}\to F_j(i\beta_j)\, \ds\frac{d \overset{\circ}\to F_j(k)}{dk}\bigg|_{k=-i\beta_j}
}.\tag 4.18$$

\noindent PROOF: With the help of (4.7) and (4.8),
 using the fact that $\overset{\circ}\to F(k)$ vanishes
 at $k=-i\beta_j$ and $F_j(k)$ vanishes at $k=i\beta_j,$ we get
$$F_j(-i\beta_j)=-2i\beta_j\, \ds\frac{d\overset{\circ}\to F(k)}{dk}\bigg|_{k=-i\beta_j},
\quad \ds\frac{dF(k)}{dk}\bigg|_{k=i\beta_j}=\ds\frac{1}{2i\beta_j}\,
\overset{\circ}\to F(i\beta_j).\tag 4.19$$
Corresponding to the Jost function $F_j(k),$ we have the scattering
matrix $S_j(k)$ defined as [4,5,11,12]
$$S_j(k):=-\ds\frac{F_j(-k)}{F_j(k)}.\tag 4.20$$
 From the first line of (2.28) of
[4] it is known that the positive constant $m_j$ is related to the residue of
the scattering matrix $S_j(k)$ at $k=i\beta_j$
via
$$m_j^2=-i\,\text{Res}[S_j(k),i\beta_j].\tag 4.21$$
It is also known [4,5,11,12] that the zero of $F_j(k)$
at $k=i\beta_j$ is simple, and hence using (4.20) in
(4.21) we obtain the first
equality in (4.17). The second equality in (4.17) follows from the use of (4.19)
in the first equality in (4.17).
 From (2.28) of [4] we have
$$g_j^2=-\ds\frac{4\beta_j^2}{F_j(-i\beta_j)^2}\,m_j^2.\tag 4.22$$
Using (4.17) in (4.22) we obtain (4.18). \qed

In Theorem~4.1, starting with the input data
 $|F(k)|$ for $k\in\bR,$ we have constructed $M$ sets
 $\{\cot\theta_j,\varphi_j(k,x)\}$ for $j=1,\dots,M,$
where
 the boundary parameter $\cot\theta_j$
and the regular solution $\varphi_j(k,x)$
are uniquely obtained via the Gel'fand-Levitan procedure
(4.11)-(4.13). Alternatively, it
is possible to get $\{\cot\theta_j,\varphi_j(k,x)\}$ for $j=1,\dots M$ via the Darboux transformation,
i.e. by using (3.1) of [4] and (3.4) of [4], via
$$\cot\theta_j=\cot\overset{\circ}\to \theta+g_j^2,\tag 4.23$$
$$\varphi_j(k,x)=\overset{\circ}\to \varphi(k,x)-
\ds\frac{g^2_j\,\overset{\circ}\to \varphi(i\beta_j,x)
\ds\int_0^x dy\,\overset{\circ}\to \varphi(k,y)\,
\overset{\circ}\to \varphi(i\beta_j,y)}{1+g^2_j\ds\int_0^x dy\,
\overset{\circ}\to \varphi(i\beta_j,y)^2},\tag 4.24$$
where $\cot\overset{\circ}\to \theta$ is the quantity
in the second equality in (4.6), $\overset{\circ}\to \varphi(k,x)$
is the quantity in the third equality in (4.6), and
$g_j$ is the Gel'fand-Levitan norming constant appearing in (4.18).

Alternatively, starting with the input data
 $|F(k)|$ for $k\in\bR,$ we can construct $M$ sets
 $\{\cot\theta_j,\varphi_j(k,x)\}$ for $j=1,\dots,M,$
where we obtain
 the boundary parameter $\cot\theta_j$
and the regular solution $\varphi_j(k,x)$
via the Marchenko procedure as follows.
First, from the input data set
$|F(k)|$ for $k\in\bR$ we obtain $\overset{\circ}\to F(k)$
as in (4.2) and obtain the set $\{\beta_1,\dots,\beta_M\}$
with the help of (4.7). Then, we use (4.8) and get
the Jost function $F_j(k)$ for each $j=1,\dots,M.$
Then, for each $\beta_j$-value we form the
scattering matrix $S_j(k)$ defined as in (4.20) and
the Marchenko norming constant $m_j$ appearing in (4.17).
We then use $S_j(k),$ $\beta_j,$ and $m_j$ to form the
Marchenko kernel
$$M_j(y):=\ds\frac{1}{2\pi}\ds\int_{-\infty}^\infty
dk\,[S_j(k)-1]\,e^{iky}+m_j^2\,e^{-\beta_j y}.$$
We next use $M_j(y)$ as input into the Marchenko integral equation
$$K_j(x,y)+M_j(x+y)+\int_x^\infty dz\,K_j(x,z)\,M_j(z+y)=0,\qquad
0\le x<y,$$
and uniquely recover $K_j(x,y).$ Then, the Jost solution
$f_j(k,x)$ is obtained via [5,9,11,12]
$$f_j(k,x)=e^{ikx}+\ds\int_x^\infty dy\,K_j(x,y)\,e^{iky},$$
and then the regular solution $\varphi_j(k,x)$ is obtained
with the help of (3.28) as
$$\varphi_j(k,x)=\ds\frac{1}{2k}\,\left[F_j(k)\,f_j(-k,x)-F_j(-k)\,f_j(k,x)\right].
\tag 4.25$$
Finally, the value of $\cot\theta_j$ is obtained
by using (4.25) in the second equality in (4.15).

Having described the recovery of all
the potentials and the bound-state parameters corresponding to
the absolute value of the Jost function,
we are now ready to describe the recovery of all candidates
for the vocal-tract radius function from the
corresponding absolute
pressure at the lips.

\noindent {\bf Theorem 4.3} {\it Assume that the vocal-tract radius $r(x)$
belongs to class $\Cal A$
and that $r(x)$ is extended beyond
$x=\ell$ as in (3.2). Let $P(k,\ell)$ be the corresponding pressure at the lips,
and let $P_\infty$ be the positive constant appearing in (2.27) and (2.28).
Let $\overset{\circ}\to F(k)$ be the quantity defined in (4.2) by using
(2.29) on the right-hand side of (4.2).
Consider also
the half-line Schr\"odinger equation
where the potential $q(x)$ is related to $r(x)$ as in (3.4)
and to the boundary condition given in (2.12) with $\cot\theta$ as in (2.13).
Let $f(k,x)$ be
the corresponding Jost solution
satisfying (3.1) and (3.3), $F(k)$ be
the Jost function given in (2.18), and $\varphi(k,x)$
be the regular solution satisfying (3.1) and (2.15)
with $\cot\theta$ as in (2.13). Then:}

\item{(a)} {\it The vocal-tract radius $r(x)$ is related to
the regular solution $\varphi(k,x)$ as in (2.16).}

\item{(b)} {\it If $r'(\ell)\ge 0,$ then the vocal-tract radius $r(x)$ for $x\in(0,\ell)$
is uniquely determined by
$|P(k,\ell)|$ known for $k>0.$ In this case the vocal-tract
radius $r(x)$ is equal to
$\overset{\circ}\to r(x)$ given by}
$$\overset{\circ}\to r(x):=\sqrt{\ds\frac{c\mu}{\pi P_\infty\,\overset{\circ}\to \varphi(0,\ell)}}
\,\overset{\circ}\to \varphi(0,x),\tag 4.26$$
{\it where $\overset{\circ}\to \varphi(k,x)$ is the regular solution given in the
third equality in (4.6) and is uniquely
obtained by the Gel'fand-Levitan procedure described in
(4.4)-(4.6) by using
the quantity $|P(k,\ell)|^2/P_\infty^2$ in place of $k^2/|F(k)|^2$
in (4.4).}

\item{(c)} {\it If $r'(\ell)<0,$ then there are $M$ candidates $r_j(x)$ for
$j=1,\dots,M$ for the vocal-tract radius $r(x)$
for $x\in(0,\ell),$ where $M$ is the maximal number of eligible resonances
related to $\overset{\circ}\to F(k)$
and is
uniquely determined by the
two conditions in (4.7) for $j=1,\dots,M.$
Each $r_j(x)$ is given by}
$$r_j(x):=\sqrt{\ds\frac{c\mu}{\pi P_\infty\,\varphi_j(0,\ell)}}\,
\varphi_j(0,x),\qquad j=1,\dots,M,\tag 4.27$$
{\it where $\varphi_j(k,x)$ is the regular solution given in the third
equality in (4.13)
and is uniquely obtained by the Gel'fand-Levitan procedure of
(4.11)-(4.13) by using the quantity
$|P(k,\ell)|^2/P_\infty^2$ in place of $k^2/|F(k)|^2$
on the right-hand side of (4.11)
and by using there the value of $g_j^2$ given in (4.18).}

\noindent PROOF: The proof of (a) follows from (2.16), and
we refer the reader to [1,2] for the proof of (2.16).
For the proof of (b) we proceed as follows. If $r'(\ell)\ge 0,$ by
Theorem~3.1(a) we know that the associated Schr\"odinger operator has no bound states,
and hence we can recover
the corresponding regular solution via the Gel'fand-Levitan procedure
as in (4.3)-(4.5) by taking into account (2.29). Once the regular solution
$\overset{\circ}\to \varphi(k,x)$ is recovered, by using (2.16) at $x=\ell$ we get
$$\ds\frac{\overset{\circ}\to r(\ell)}{\overset{\circ}\to r(0)}=\overset{\circ}\to \varphi(0,\ell).\tag 4.28$$
 From (2.28) we have
$$\overset{\circ}\to r(0)\,\overset{\circ}\to r(\ell)=\ds\frac{c\mu}{\pi P_\infty}.\tag 4.29$$
Using (4.28) and (4.29) we obtain
$$\overset{\circ}\to r(0)=\sqrt{\ds\frac{c\mu}{\pi P_\infty\,\overset{\circ}\to \varphi(0,\ell)}}.\tag 4.30$$
Finally, using (4.30) in (2.16) we extract the value of $\overset{\circ}\to r(x)$ as in
(4.26). Thus, we have proved (b). Let us now prove (c).
If $r'(\ell)<0,$ by
Theorem~3.1(a) we know that the associated Schr\"odinger operator has
exactly one bound state and from Theorem~4.1 we know that there
are precisely $M$ ways to choose the bound state at $k=i\beta_j$
for $j=1,\dots,M.$ The choice $k=i\beta_j$ yields the Jost function
$F_j(k)$ given in (4.8) with the Gel'fand-Levitan norming constant $g_j$ appearing
in (4.10) or equivalently in (4.18). The corresponding regular solution is obtained
as in the third equality in (4.13)
via the
Gel'fand-Levitan procedure described in (4.11)-(4.13). The corresponding vocal-tract
radius is recovered from the regular solution as in
(4.27), by proceeding as in the proof of (b) leading to (4.26).
Thus, the proof of (c)
is complete. \qed

In Theorem~4.3, we have seen that the absolute pressure at the lips uniquely
determines the vocal-tract radius function in the absence of bound states for
the corresponding Schr\"odinger operator, and we have used
$\overset{\circ}\to r(x)$ to denote that radius function. In the presence of a bound state,
we have seen that there are $M$ candidates for the vocal-tract radius function,
and we have used $r_j(x)$ to denote the choices for $j=1,\dots,M,$
where $M$ is the number appearing in Theorem~4.1, i.e. the number
of positive $\beta_j$-values satisfying (4.7).
We next show that we have
$r_j(x)\not\equiv \overset{\circ}\to r(x)$ for $j=1,\dots,M,$ and also
$r_j(x)\not\equiv r_s(x)$ if $1\le j<s\le M.$

\noindent {\bf Proposition 4.4} {\it Let
the vocal-tract radius $r(x)$ belong to class $\Cal A,$ and
consider the inverse problem of
recovery of $r(x)$
for $x\in(0,\ell)$ from the absolute pressure $|P(k,\ell)|$
for $k>0$ measured at the lips. Let $\overset{\circ}\to r(x)$ be the
quantity in (4.26) and let $r_j(x)$ for $
j=1,\dots,M$ be the quantity given in (4.27)
associated with the corresponding positive constant $\beta_j$ appearing in
the proof of Theorem~4.3.}

\item{(a)} {\it If $M\ge 1$ then we must have}
$$r_j(x)\not\equiv \overset{\circ}\to r(x),\qquad j=1,\dots,M.\tag 4.31$$

\item{(b)} {\it If $M\ge 2$ then we must have}
$$r_j(x)\not\equiv r_s(x),\qquad 1\le j<s\le M.\tag 4.32$$

\noindent PROOF: By Proposition~3.3(b) and Theorem~4.3(b) we know that $\overset{\circ}\to r(x)$
is associated with the Schr\"odinger operator having no bound states
and hence $\overset{\circ}\to r'(\ell)\ge 0.$
On the other hand, each $r_j(x)$ corresponds to one bound state for the associated Schr\"odinger equation
and hence $r_j'(\ell)<0.$ Thus, we must have (4.31). An alternate proof of (4.31)
can be given as follows. If we had
$r_j(x)\equiv \overset{\circ}\to r(x),$ we would also have
$r_j(0)=\overset{\circ}\to r(0)$ and $r'_j(0)=\overset{\circ}\to r'(0).$ However,
because of (2.13) and (4.23)
we would have
$$-\ds\frac{r'_j(0)}{r_j(0)}=-\ds\frac{\overset{\circ}\to r'(0)}{\overset{\circ}\to r(0)}+g_j^2,$$
implying $g_j^2=0.$ However, the Gel'fand-Levitan norming constant $g_j$
appearing in (4.1) must be positive, yielding a contradiction. Thus, (4.31)
must hold. Let us now turn to the proof of (b). If (4.32) were not true, i.e.
if we had $r_j(x)\equiv r_s(x)$ with $\beta_j\ne \beta_s,$ then because of (3.4) the corresponding potentials
$q_j(x)$ and $q_s(x)$ would coincide and because of (2.13) the corresponding
boundary parameters $\cot\theta_j$ and $\cot\theta_s$ would also coincide. However, this
would force the corresponding Jost functions $F_j(k)$ and $F_s(k)$ to coincide as well.
Because of (4.8) and the fact that $\overset{\circ}\to F(k)$ given in
(4.2) does not vanish
in $\bCp,$ as assured by Theorem~3.1(b), we would then have $\beta_j=\beta_s,$
which is a contradiction. Thus, (4.32) must hold. \qed

Let us remark that the $M$ quantities $r_j(x)$
constructed as in (4.27) are not all necessarily admissible as vocal-tract radii.
The admissibility is satisfied by those $r_j(x)$ for which
$r_j(x)>0$ for $x\in [0,\ell].$ Those $r_j(x)$ with a zero in $x\in[0,\ell]$
are inadmissible.
We can equivalently state the admissibility
as $r_j(\ell)>0.$ The necessity for $r_j(\ell)>0$ for the admissibility is clear.
The sufficiency can be seen as follows.
Since we already have the continuity of $r_j(x)$ on $x\in(0,\ell)$
 and we already know that $r_j(0)>0,$ having $r_j(\ell)>0$ indicates
that the number of zeros of $r_j(x)$ including multiplicities
in the interval $(0,\ell)$ must be either zero or an even integer.
With the help of (3.10) we then conclude that the number of zeros of
the corresponding
$\varphi_j(0,x)$ in $(0,\ell)$ must be either zero or
an even integer. On the other hand,
by Theorem~3.1(c), we know that the number of zeros of
$\varphi_j(0,x)$ in $(0,+\infty)$ is either zero or one.
Thus, we have proved that $r_j(x)>0$ for $x\in [0,\ell]$
when $r_j(\ell)>0.$

In the next proposition, we provide an equivalent admissibility criterion for
$r_j(x)$ constructed as in (4.27).

\noindent {\bf Proposition 4.5} {\it Let the absolute pressure
at the lips, $|P(k,\ell)|,$ correspond through (2.27) and (2.29)
to the absolute value
of the Jost function
$|F(k)|$ having exactly one bound state, which may occur at
$k$-values given by
$k=i\beta_j$ for $j=1,\dots,M$ for some integer $M$
with $M\ge 1$ as indicated in Theorem~4.1(b). For each value
 of $j,$ let $\varphi_j(k,x)$ be the regular solution appearing in the third
equality in (4.13), $r_j(x)$ be the quantity in (4.27), and
$\overset{\circ}\to \varphi(k,x)$ be the regular solution appearing
in the third equality in (4.6). Then, $r_j(x)$ is admissible as a vocal-tract radius
if and only if we have}
$$2\beta_j \ds\int_0^\ell dy\,\overset{\circ}\to \varphi(0,y)\,\overset{\circ}\to \varphi(i\beta_j,y)<\overset{\circ}\to \varphi(0,\ell)\,\overset{\circ}\to \varphi(i\beta_j,\ell).\tag 4.33$$

\noindent PROOF: It is enough to prove that $r_j(\ell)>0$ if and only if (4.33)
holds. Since $r_j(0)>0,$ from (3.10) it follows that $r_j(\ell)>0$
is equivalent to $\varphi_j(0,\ell)>0.$ Evaluating (4.24) at $k=0$ and $x=\ell$ we
obtain
$$\varphi_j(0,\ell)=\overset{\circ}\to \varphi(0,\ell)-
\ds\frac{g^2_j\,\overset{\circ}\to \varphi(i\beta_j,\ell)
\ds\int_0^\ell dy\,\overset{\circ}\to \varphi(0,y)\,
\overset{\circ}\to \varphi(i\beta_j,y)}{1+g^2_j\ds\int_0^\ell dy\,
\overset{\circ}\to \varphi(i\beta_j,y)^2}.\tag 4.34$$
 From the positivity of $g_j^2,$
we are guaranteed that the denominator in (4.10) is positive.
We replace $g_j^2$ on the right-hand side of (4.34)
by the right-hand side of (4.10), and we see that $\varphi_j(0,\ell)>0$
if and only if we have
$$\overset{\circ}\to \varphi(0,\ell)>\ds\frac{2\beta_j\,\overset{\circ}\to \varphi(i\beta_j,\ell)\ds\int_0^\ell dy\,\overset{\circ}\to \varphi(0,y)\,
\overset{\circ}\to \varphi(i\beta_j,y)}
{\overset{\circ}\to \varphi(i\beta_j,\ell)^2}.\tag 4.35$$
Furthermore, both $\overset{\circ}\to\varphi(0,x)$ and $\overset{\circ}\to\varphi(i\beta_j,x)$
are positive in $x\in[0,+\infty)$ because each quantity is equal to
one by the first equality in (4.3) and neither quantity has any zeros
in $x\in(0,+\infty).$ The absence of zeros follows from the fact [10,20] that
the number of zeros for each quantity would be either equal to
or one less than the
number of bound states, and the number of bound states
associated with $\overset{\circ}\to \varphi(k,x)$ is zero.
Thus, (4.35) yields (4.33). \qed

In the solution of our inverse problem, we have
assumed that the length $\ell$ of the vocal
tract is known. Let us briefly comment on the case if $\ell$
is not known.
When the conditions in Theorem~3.4 are satisfied by the vocal-tract radius, from (3.36)
we see that the value of $\ell$ can be determined from the large-$k$ asymptotics of
the absolute pressure if $r''(\ell^-)\ne 0.$ Alternatively, if we do not know the
value of $\ell,$ we can solve our inverse problem and recover
$r(x)$ in the larger interval $x\in(0,\tilde \ell)$ for some
$\tilde \ell\ge \ell$ and estimate the actual $\ell$-value as
the smallest $x$-value beyond which
$r''(x)/r(0)$ becomes zero.

\vskip 10 pt
\noindent {\bf 5. AN ALGORITHM FOR THE SOLUTION OF THE INVERSE PROBLEM}
\vskip 3 pt

In Section~4 we have considered the inverse problem of
recovery of the vocal-tract radius $r(x)$ for $x\in(0,\ell)$ from
the absolute pressure at the lips, i.e. from the
input data set $|P(k,\ell)|$ for $k>0.$ We have seen that we have the
unique recovery if $r'(\ell)\ge 0$ and we have
up to an $M$-fold
nonuniqueness in the recovery if $r'(\ell)<0,$
where $M$ is the nonnegative integer equal to the
number of positive $\beta_j$-values at which the two
conditions in (4.7) are satisfied.

In this section, when $r'(\ell)\ge 0,$
we provide an algorithm to uniquely recover
the vocal-tract radius from the absolute pressure. In this
algorithm, we use the quantity $B(t)$ as input related to the absolute pressure
as in (5.1), where $t$ can be interpreted as time, and hence
we call our algorithm a time-domain algorithm.
This algorithm
consists of the following steps:

\item{(a)} From the input data set
$|P(k,\ell)|$ for $k>0,$ determine $P_\infty$ via (2.27).

\item{(b)} Define $B(t)$ for $0< t<2 \ell$ as
$$B(t):=\ds\frac{2}{\pi}\ds\int_0^\infty dk\,
\left[\ds\frac{|P(k,\ell)|^2}{P_\infty^2}-1\right]
\left(\cos kt\right).\tag 5.1$$
With the help of (2.29) we observe that $B(t)=\overset{\circ}\to G(0,t),$
where $\overset{\circ}\to G(x,y)$ is the Gel'fand-Levitan kernel appearing in (4.4).

\item{(c)} Use $B(t)$ as input to the overdetermined system
$$\cases [r(x)^2\,w_x(x,t)]_x-r(x)^2\,w_{tt}(x,t)=0,\qquad 0< x<t<2\ell-x,\\
\stretch
w_t(0,t)=B(t),\qquad 0<t<2\ell,\\
\stretch
w_x(0,t)=0,\qquad 0<t<2\ell,\\
\stretch
w(x,x)=\ds\frac{r(0)}{r(x)}, \qquad 0< x<\ell.\endcases
\tag 5.2$$

\item{(d)} Recover $r(x)/r(0)$ uniquely from the
overdetermined system in (5.2). This can be done numerically by using, for example, the
downward-continuation scheme based on (3.5.9a)-(3.5.11) of [6].

\item{(e)} Having $r(x)/r(0)$ at hand for $x\in(0,\ell),$ retrieve the
vocal-tract radius $r(x)$
as
$$r(x)=\left[\ds\frac{r(x)}{r(0)}\right]\ds\sqrt{\ds\frac{c\mu}{\pi P_\infty}}
\left[\ds\frac{1}{\ds\sqrt{r(\ell)/r(0)}}\right],\qquad x\in(0,\ell).\tag 5.3$$

In the following theorem, we elaborate on our time-domain
algorithm summarized above. We show that the vocal-tract radius
$r(x)$ and the absolute pressure $|P(k,\ell)|$ are
related to each other through (5.1) and the overdetermined
system (5.2). We prove that
the system in (5.2) is uniquely solvable for
$r(x)/r(0)$ if
$B(t)$ is already associated with
a vocal-tract radius $r(x)$ in class $\Cal A$ with
$r'(\ell)\ge 0.$ We also provide
a justification for the expression for $r(x)$ given in (5.3).
We remark that our theorem is valid
only when $r'(\ell)\ge 0,$ i.e. when we have
either the first or second possibility in Theorem~3.1(c)
but not the third or fourth possibility.

\noindent {\bf Theorem 5.1} {\it Assume that $r(x)$ belongs to class
$\Cal A$ and that $r'(\ell)\ge 0.$ Let $|P(k,\ell)|$
be the corresponding absolute pressure at the lips, as given in (2.23).
Define $B(t)$ as in
(5.1) and use it as input into the system (5.2). Then, the overdetermined system (5.2)
is uniquely solvable for $r(x)/r(0)$ when we use
$|P(k,\ell)|$ as input through $B(t).$ Furthermore, $r(x)$ is obtained from
$r(x)/r(0)$ as in (5.3).}

\noindent PROOF: Let us define
$$u(x,t):=\ds\frac{1}{2\pi} \ds\int_{-\infty}^\infty dk\,\ds\frac{k\,f(k,x)}{F(k)}
\,e^{-ikt},\tag 5.4$$
where $f(k,x)$ is the Jost solution to
(3.1) appearing in (3.3) and $F(k)$ is the Jost
function appearing in (2.18). As indicated in Theorem~3.1(b),
the assumption $r'(\ell)\ge 0$ guarantees that
$F(k)$ does not have any zeros in $\bCpb,$ except perhaps for a simple zero at
$k=0.$ Thus, the integrand in (5.4) does not
have any poles in $k\in\bCpb.$
One can directly verify that $u(x,t)$ is
the unique solution to the
system
$$\cases u_{xx}(x,t)-u_{tt}(x,t)=q(x)\,u(x,t),\qquad 0< x<t<2\ell-x,\\
\stretch
u(0,t)=B(t),\qquad 0< t<2 \ell,\\
\stretch
u_x(0,t)=-(\cot\theta)\,B(t),\qquad 0< t<2 \ell,\\
\stretch
u(x,x)=\cot\theta-\ds\frac{1}{2}\,\int_0^x ds\,q(s), \qquad 0< x<\ell,\endcases\tag 5.5$$
where $q(x)$ is the quantity given in (3.4) and
$\cot\theta$ is given by (2.13). The verification of the first
line in (5.5) can be achieved by taking the appropriate derivatives
of the right-hand side of (5.4) and by using the fact that
$f(k,x)$ satisfies (3.1). In order to verify the second line in
(5.5), let us write (5.4) as
$$u(x,t)=\delta(x-t)+\ds\frac{1}{2\pi} \ds\int_{-\infty}^\infty dk\,
Q(k,x)
\,e^{ik(x-t)},\tag 5.6$$
where we have used the fact that the Dirac delta distribution is given by
$$\delta(t)=\ds\frac{1}{2\pi} \ds\int_{-\infty}^\infty dk\,e^{ikt},\tag 5.7$$
and we have defined
$$Q(k,x):=\ds\frac{k\,m(k,x)}{F(k)}-1,\tag 5.8$$
with $m(k,x)$ being the quantity defined in (3.37).
For each fixed $x\in[0,\ell]$ we already know [5,11,12]
that $m(k,x)$ is analytic in $k\in\bCp,$ continuous in $k\in\bCpb,$ and
behaves like
$1+O(1/k)$ as $k\to\infty$ in $\bCpb.$ In fact, we have
$$m(k,x)=1-\ds\frac{1}{2ik} \ds\int_x^\infty dy\,
q(y)+\ds\frac{1}{2ik} \ds\int_x^\infty dy\,
q(y)\,e^{2ik(y-x)}+O\left(\ds\frac{1}{k^2}\right),\qquad
k\to\infty \text{ in } \bCpb.\tag 5.9$$
The large-$k$ asymptotics of the Jost function is given by [5]
$$F(k)=k-i\,\cot\theta+\ds\frac{i}{2}\ds\int_0^\infty dy\,
q(y)+\ds\frac{i}{2}\ds\int_0^\infty dy\,
q(y)\,e^{2iky}+O\left(\ds\frac{1}{k}\right),\qquad
k\to\infty \text{ in } \bCpb.\tag 5.10$$
Using (5.9) and (5.10) in (5.8), as $k\to\infty$ in
 $\bCpb$ we obtain
$$\aligned Q(k,x)=&\ds\frac{i}{k}\left[\cot\theta-\ds\frac{1}{2}\ds\int_0^x dy\,
q(y)\right]\\ &
-\ds\frac{i}{2k}\left[\ds\int_0^\infty dy\,
q(y)\,e^{2iky}+\ds\int_x^\infty dy\,
q(y)\,e^{2ik(y-x)}\right]+O\left(\ds\frac{1}{k^2}\right).\endaligned\tag 5.11$$
As a result of $r'(\ell)\ge 0,$
we already know by Theorem~3.1(b) that
$k/F(k)$ is analytic
in $k\in\bCp$ and continuous in $\bCpb.$
Thus, for each fixed $x\in[0,+\infty),$ the quantity $Q(k,x)$
is analytic
in $k\in\bCp,$ continuous in $k\in\bCpb,$ and $O(1/k)$ as $k\to\infty$ in
$\bCpb.$
Note that
$$Q(-k,0)=Q(k,0)^\ast,\qquad k\in\bR,\tag 5.12$$
as a result of
(2.20) and that
[5] $f(-k,x)=f(k,x)^\ast$ for $k\in\bR.$
Using $x=0$ and $t>0$ in (5.4), we obtain
$$u(0,t)=\ds\frac{1}{2\pi} \ds\int_{-\infty}^\infty dk\,Q(k,0)\,e^{-ikt}.\tag 5.13$$
 From (5.12) and the facts that
$Q(k,0)$ is analytic in $\bCp,$ continuous in $\bCpb,$ and
$O(1/k)$ as $k\to\infty$ in $\bCpb,$ it follows that the right-hand side of (5.8) can be evaluated [14]
in terms of the real part of $Q(k,0)$ as
$$\ds\frac{1}{2\pi} \ds\int_{-\infty}^\infty dk\,Q(k,0)\,e^{-ikt}=
\ds\frac{2}{\pi}\int_0^\infty dk\, \text{Re}[Q(k,0)]\,\cos(kt).\tag 5.14$$
One can evaluate $\text{Re}[Q(k,0)]$ with the help of (2.17) and (5.14), or simply by using
Proposition~3.6 of [5], and
we have
$$\text{Re}[Q(k,0)]=\ds\frac{k^2}{|F(k)|^2}-1,\qquad k\in\bR.\tag 5.15$$
With the help of (2.29), (5.1), and (5.13)-(5.15), we see that
the second line of (5.5) is satisfied. Let us now prove that
the third line of (5.5) is satisfied. From (5.4) we get
$$u_x(0,t)=\ds\frac{1}{2\pi} \ds\int_{-\infty}^\infty dk\,\ds\frac{k\,f'(k,0)}{F(k)}
\,e^{-ikt}.\tag 5.16$$
Using (2.17) we can replace $f'(k,0)$ in (5.16) by
$i F(k)-(\cot\theta) f(k,0),$ yielding
$$u_x(0,t)=\ds\frac{1}{2\pi} \ds\int_{-\infty}^\infty dk\,ik\,e^{-ikt}-
\ds\frac{\cot\theta}{2\pi} \ds\int_{-\infty}^\infty dk\,\ds\frac{k\,f(k,0)}{F(k)}
\,e^{-ikt}.\tag 5.17$$
With the help of (5.7) we recognize the first term on the right-hand side
of (5.17) as
$-\delta'(t),$ which is zero because
we have $t>0.$ With the help of (5.6)
we see that the second term on the right-hand side of (5.17) is
equal to $-(\cot\theta)\, u(0,t),$ and hence the third line
of (5.5) follows from the second line of (5.5).
Let us now prove that the fourth line
of (5.5) holds. First, we remark that
by $u(x,x)$ on the fourth line of (5.5), we actually mean
$u(x,x^+).$ In fact, we conclude $u(x,x^-)=0$ from (5.4)
as a result of the facts that
$Q(k,x)$ is analytic in $k\in\bCp,$ continuous in $k\in\bCpb,$ and
$O(1/k)$ as $k\to\infty$ in $\bCpb.$
In the evaluation of
the difference $u(x,x^+)-u(x,x^-)$ with the help of
(5.6) and (5.11) we see that the only nonzero contribution comes
 from the first term on the right-hand side in (5.11).
With the help of
$$\ds\frac{1}{2\pi}  \ds\int_{-\infty}^\infty dk\,
\ds\frac{e^{ik\alpha}}{k+i0^+}=-i\Theta(-\alpha),$$
where $\Theta(\alpha)$ is the Heaviside function
such that $\Theta(\alpha)=1$ when $\alpha>0$ and
$\Theta(\alpha)=0$ when $\alpha<0,$
we obtain
$$u(x,x^+)=\ds\frac{i}{2\pi}  \ds\int_{-\infty}^\infty \ds\frac{dk\, e^{-ik 0^+}}{k+i 0^+}\,
\left[\cot\theta-\ds\frac{1}{2}\ds\int_0^x dy\,
q(y)\right],$$
confirming the fourth line of (5.5).
Next, letting
$$w(x,t):=\ds\frac{r(0)}{r(x)}\,\left[1+\ds\int_x^t ds\,u(x,s)\right],$$
one can prove, by using elementary calculus rules, that (5.2) is satisfied by
$w(x,t),$ $r(x),$ and $B(t).$
Then, Theorem~2 of [19] implies that $r(x)/r(0)$ appearing in the last line
of (5.2) is uniquely determined. This is done by showing that
the map $B(t)\mapsto r(x)/r(0)$
is Lipschitz continuous from the linear space $L^2(0,2T)$ to
the Sobolev space
$H^1(0,T)$ for any finite positive $T.$
Having $r(x)/r(0)$ uniquely determined, with the help of
(2.16) and (4.27), we obtain (5.3). \qed

\vskip 10 pt
\noindent {\bf 6. EXAMPLES}
\vskip 3 pt

In this section through some examples
we illustrate the inverse problem of recovery of the
vocal-tract radius $r(x)$ for $x\in(0,\ell)$ from the absolute pressure $|P(k,\ell)|$
known for $k\in(0,+\infty).$

In the first example below we consider a linear vocal-tract radius function.
We show that two such linear functions, one with positive slope and
the other with a negative slope, correspond to the same absolute pressure.

\noindent{\bf Example 6.1} Let us assume that the absolute pressure
at the lips is given by
$$|P(k,\ell)|=\ds\frac{k\,P_\infty}{\sqrt{k^2+a^2}},\qquad k\in(0,+\infty),\tag 6.1$$
where $a$ is a positive constant and $P_\infty$ is the asymptotic value of the absolute
pressure given in (2.27). With the help of (2.29) we find that
the corresponding Jost function
$\overset{\circ}\to F(k)$ with no bound states
appearing in (4.26) and the Jost function
$F_1(k)$ with one bound state appearing in (4.8) are respectively given by
$$\overset{\circ}\to F(k)=k+i\,a,\quad F_1(k)=k-ia,\tag 6.2$$
and hence we have $M=1,$ i.e.
the maximal number of eligible resonances is one.
We note that $\overset{\circ}\to F(k)$ given in (6.2) corresponds to the zero potential
$\overset{\circ}\to q(x)\equiv 0$ and to the boundary parameter
$\cot \overset{\circ}\to \theta=-a,$
yielding the vocal-tract radius
$$\overset{\circ}\to r(x)=\sqrt{\ds\frac{c\mu}{\pi P_\infty (1+a \ell)}}\,(1+ax),
\qquad x\in(0,\ell),\tag 6.3$$
which agrees with (4.26). On the other hand,
the Jost function $F_1(k)$ appearing in (6.2) corresponds to the zero potential
$q_1(x)\equiv 0,$ the boundary parameter $\cot\theta_1=a,$ and
the vocal-tract radius $r_1(x)$ given by
$$r_1(x)=\sqrt{\ds\frac{c\mu}{\pi P_\infty (1-a \ell)}}\,(1-ax),
\qquad x\in(0,\ell),\tag 6.4$$
which agrees with (4.27).
Thus, in this example, both $\overset{\circ}\to r(x)$ and
$r_1(x)$ correspond to the data set consisting of the absolute pressure $|P(k,\ell)|$ given in (6.1).
We note that $\overset{\circ}\to r'(\ell)/\overset{\circ}\to r(0)=a$ and
$r_1'(\ell)/r_1(0)=-a,$ and hence if we include in our input data set the sign of
the derivative of the vocal-tract radius at the lips then we can uniquely
determine the vocal-tract radius. In this example, if $a\ge 1/\ell$ then
$r_1(x)$ does not belong to class $\Cal A,$ and hence in this case the only acceptable
vocal-tract radius is given by (6.3). On the other hand, if $a\in(0,1/\ell),$
then both $\overset{\circ}\to r(x)$ and $r_1(x)$ belong to class $\Cal A$ and
correspond to the absolute pressure given in (6.1). Let us remark that
$\overset{\circ}\to r(x)$ given in (6.3) corresponds to (3.5) and that $r_1(x)$
given in (6.4) corresponds to (3.7).
In this example, if $a=0$ and hence if we have
$|P(k,\ell)|\equiv P_\infty,$ then we get
$\overset{\circ}\to r(x)=\sqrt{c\mu/(\pi P_\infty)},$ which
corresponds to a uniform vocal tract.

In the next example, we illustrate the solution of the direct problem numerically
by using the alternate method outlined following Theorem~2.2.
Note that in Examples~6.2 and 6.3 we use the vocal-tract area $A(x)$ instead of the radius $r(x)$ because these two examples concern the data set from [18] specified
in terms
of $A(x),$ not in terms of $r(x).$
Since $r(x)$ and $A(x)$ are related to each other as in (1.1), one
can easily describe the results in these two examples also in terms of $r(x).$

\noindent{\bf Example 6.2} In this example we use the data
for the vowel /\ae/ (the vowel in the word ``cat") given in
the sixth column of Table 1 in [18].
The data set consists of 44 numbers corresponding to the MRI measurements of the
vocal-tract cross-sectional area of the author
of [18] in 2002. The length of the vocal tract is $\ell=16.11$ cm.
The 44 data points are indicated by an asterisk in the first plot of
Figure~6.1. We have used a linear interpolation to obtain the cross-sectional area $A(x)$
as a continuous function of $x,$ but we do not display
 the interpolated $A(x)$ in Figure~6.1.
 Due to the fairly large number of sampling points, i.e. 44 points, the interpolated $A(x),$
 if plotted, looks fairly smooth. We remark that
the continuous curve in the first plot is not the interpolated $A(x).$
In the example, we use $c=34300$ cm/sec for the sound speed and
$\mu=0.0012$ gm/cm$^3$ for the air density. We then use the four steps (a)-(d)
listed following Theorem~2.2
in order to obtain $|P(k,\ell)|,$ the absolute pressure at the lips, as a function of
$k.$ In solving the system (2.2) for $\tilde P(k,x)$ and
$\tilde V(k,x)$ with the initial values given in (2.26), we have
employed the standard MATLAB ODE solver {\tt{ode45}} using
$k_j=j\,\Delta k$ with $j=1,\dots, 1000$ and $\Delta k=0.003.$
The result is indicated in the second plot
in Figure~6.1, where the graph of $|P(k,\ell)|$ is shown for $k\in(0,3).$

In the next example, we illustrate the solution of the inverse problem numerically
by using the absolute pressure data at the lips given in Example~6.2. As seen from
the first discrete-data plot in Figure~6.1, for the vowel /\ae/ we have $r'(\ell)>0,$
and hence the use of the time-domain method of Section~5
is justified.

\noindent{\bf Example 6.3} In this example, we use the absolute pressure evaluated
in Example~6.2 from the vocal-tract cross-sectional area data from [18]. We use the
time-domain method of Section~5 to reconstruct the corresponding cross-sectional area
and compare the computed $A(x)$ with the values of $A(x)$ used as input in Example~6.2.
We have plotted the computed $A(x)$ as a continuous curve in the first plot of
Figure~6.1, and we recall that the values of $A(x)$ used as input in Example~6.2 are
shown as a discrete-data plot also in the first plot of
Figure~6.1, indicating a fairly accurate reconstruction. We use the following steps in
the reconstruction of $A(x)$ from the absolute pressure.

\item {(a)} Our data set consists of the finite sequence of 1000 points
corresponding to the
values of $|P(k_j,\ell)|$ with $k_j=0.003 j$ for $j=1,\dots,1000.$ We use $|P(k_{1000},\ell)|$
as $P_\infty.$ We could alternatively estimate the value of $P_\infty$ by averaging the tail of our
finite sequence of the absolute pressure values, which could improve our reconstruction.

\item {(b)} In computing a discrete version of $B(t)$ given in (5.1), we use
the discrete Fourier cosine transform. We also take into account the asymptotic behavior
of $|P(k,\ell)|,$ which is given
    in (3.36). We ignore the oscillations in the tail of $|P(k,\ell)|$ given
    in the second plot of Figure~6.1,
    and hence in (3.36) we assume that $r''(\ell^-)=0.$ In fact, we use the approximation
    $$\frac {|P(k,\ell)|^2}{|P_{\infty}|^2}=1+\frac {C}{k^2},\qquad k\ge 3,\tag 6.5$$
where we estimate the value of the constant $C$ in (6.5) by using $k=k_{1000},$ or
equivalently $k=3.$ We could alternatively estimate $C$ by using the tail of our finite data sequence,
yielding a more accurate evaluation of $B(t).$

\item {(c)} We numerically solve the overdetermined boundary-value problem (5.2)
with the input $B(t)$ estimated in the previous step. For this purpose we use a MATLAB routine based on the downward-continuation algorithm described in (3.5.9a)-(3.5.11) of [6].
We thus obtain a numerical approximation of $A(x)/A(0)$ for $x\in(0,\ell).$ Here, we assume that
we know the value of $\ell$ as $\ell=16.11$ cm.

\item {(d)} Having the numerical values
of $A(x)/A(0)$ at hand,
we compute the values of $A(x)$ for $x\in(0,\ell)$ by using the equivalent of (5.3) given by
 $$A(x)=\left[\ds\frac{A(x)}{A(0)}\right]\left[\ds\frac{c\mu}{P_\infty}\right]
\left[\ds\frac{1}{A(\ell)/A(0)}\right],\qquad x\in(0,\ell),$$
    with $c=34300$ cm/sec and $\mu=0.0012$ gm/cm$^3.$
    We have plotted the computed value of $A(x)$ in the first plot of Figure~6.1, which
     looks like a continuous curve as a result of the large number of $A(x_j)$-values
     computed.
    If we did not know the value of $\ell,$ we could first evaluate
    $A(x)/A(0)$ in the larger interval $x\in(0,\tilde \ell)$ with
    a fairly large value of $\tilde\ell$
    and estimate the actual
    $\ell$-value beyond which $r''(x)/r(0)$ becomes zero,
    where $r(x)$ is related to $A(x)$ as in (1.1).

\vskip 10 pt

\centerline{\hbox{
\psfig{figure=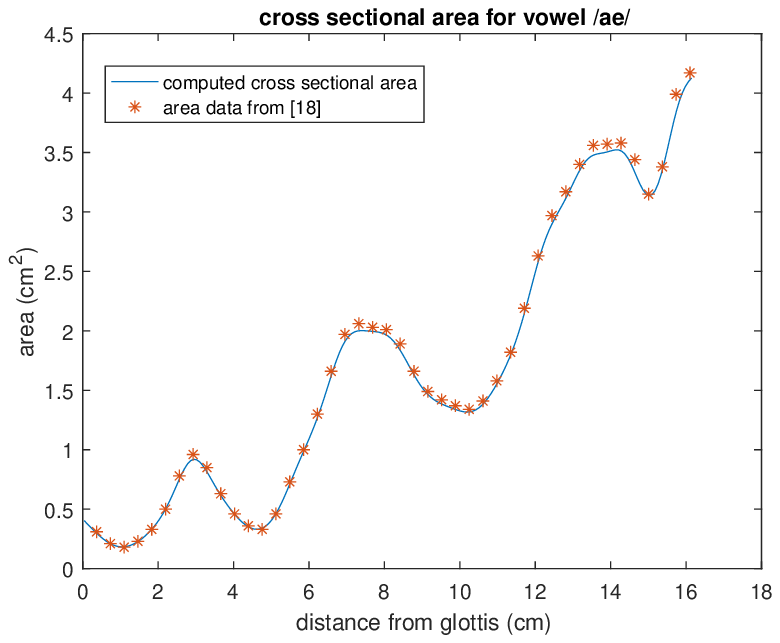,width=3 truein,height=2 truein}\qquad
\psfig{figure=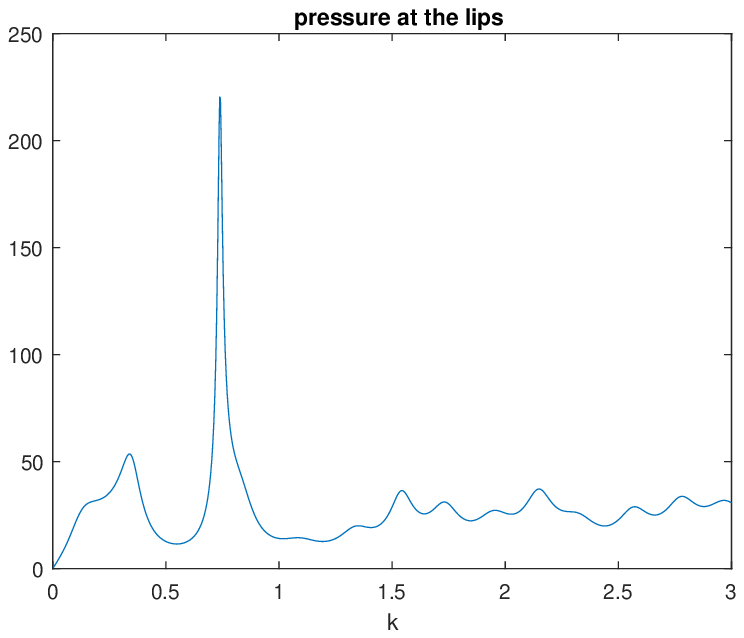,width=3 truein,height=2 truein}}}

\centerline{{\bf Figure~6.1}
$A(x)$ for /\ae/ and
$|P(k,\ell)|$ at the lips, respectively, in Examples~6.2
and 6.3.}

In the next example, we discuss some vocal-tract radius functions related to the half-line Schr\"odinger equation with the boundary condition as in (2.12) and (2.13) and
when the corresponding potential is compactly supported in $x\in[0,\ell]$ with a constant
value in that interval.

\noindent {\bf Example 6.4} Let us assume that the potential $q(x)$ given in (3.4)
vanishes when $x>\ell,$ where $\ell=16$ cm and it has the constant value $q(x)=v$ on
$x\in(0,\ell).$ Let us also assume that the value of $\cot\theta$ is related to $r(x)$ as in (2.13). Once we specify $v$ and $\cot\theta,$ we know that the
corresponding regular solution $\varphi(k,x)$ is uniquely determined. In fact,
the corresponding Schr\"odinger equation is exactly solvable, and
we are able to obtain the regular solution $\varphi(k,x),$ the Jost function $F(k),$
and any other relevant quantities by proceeding as in Example~6.2 of [4].
Let us remark that from (4.27) and (4.30)
we see that if we specify the value of $r(0)$ then the value of $P_\infty$ is uniquely
determined, and conversely if we specify the value of $P_\infty$ then the value
of $r(0)$ is uniquely determined.

\item{(a)} For example, if we choose $r(0)=1/10,$
$v=1/200,$ $\cot\theta=-1,$ then we get
$P_\infty=61.366\overline{5},$ no bound states, and
no eligible resonances. The corresponding absolute pressure at the lips given
in (2.23) and the
vocal-tract radius given in (4.26) are shown in Figure~6.2.
We remark that we use an overline on a digit to indicate a round off.

\vskip 10 pt

\centerline{\hbox{\psfig{figure=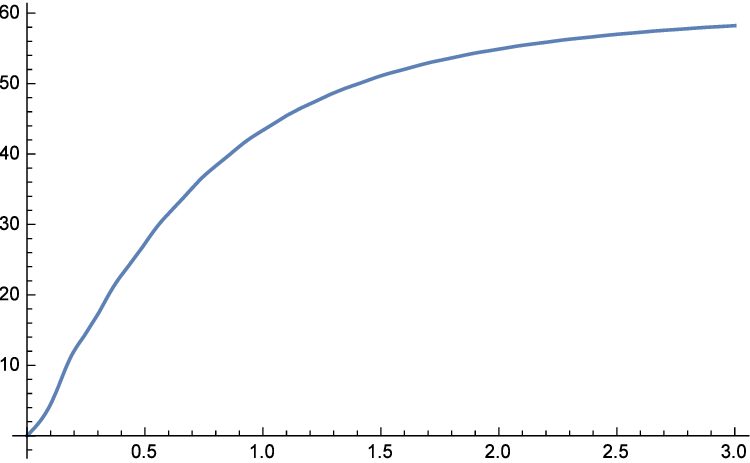,width=2 truein,height=2 truein}\qquad
\psfig{figure=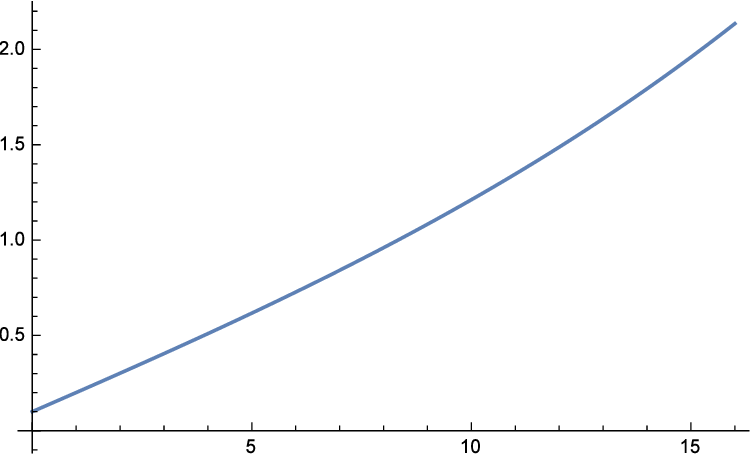,width=2 truein,height=2 truein}}}

\centerline{{\bf Figure~6.2}
$|P(k,\ell)|$ vs $k$ and $\overset{\circ}\to r(x)$ vs $x,$
respectively, in Example~6.4(a).}

\item{(b)}
If we choose $P_\infty=60,$
$v=-1/100,$ $\cot\theta=1/100,$ then there is exactly one bound state at
$k=i\beta_1,$ with $\beta_1=0.085367\overline{2}$
and the Gel'fand-Levitan norming constant
 $g_1=0.28256\overline{5},$ and there are
 no eligible resonances. In other words, in this specific case
we have $N=1$ and $M=1.$ However, in this case the quantity $r_1(x)$
given in (4.27) is inadmissible because it yields $r_1(\ell)<0.$
The corresponding absolute pressure $|P(k,\ell)|$ given in
(2.23) and the corresponding
vocal-tract radius $\overset{\circ}\to r(x)$ given in (4.26) are shown in Figure~6.3.
In the evaluation of $\overset{\circ}\to r(x)$ we have used the Darboux transformation formula
given in (3.7) of~[4].

\vskip 10 pt

\centerline{\hbox{\psfig{figure=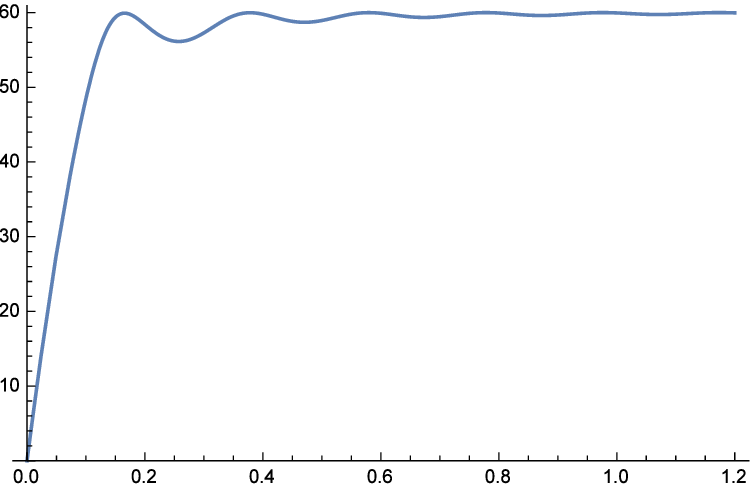,width=2 truein,height=2 truein}\qquad
\psfig{figure=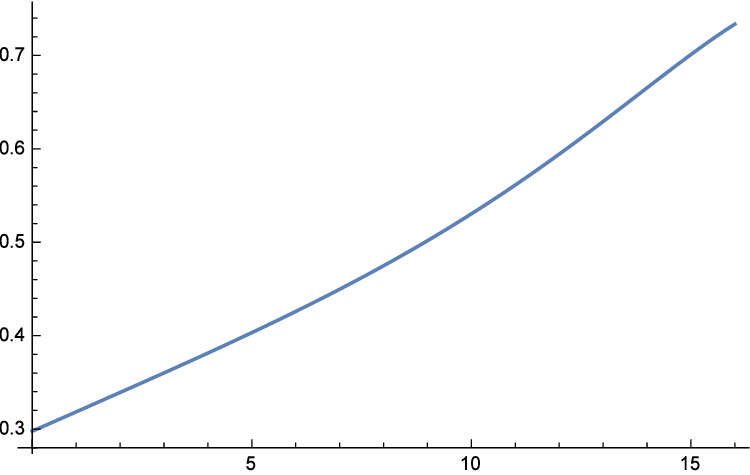,width=2 truein,height=2 truein}}}

\centerline{{\bf Figure~6.3}
$|P(k,\ell)|$ vs $k$ and $\overset{\circ}\to r(x)$ vs $x,$
respectively, in Example~6.4(b).}

\item{(c)} If we choose $P_\infty=60,$ $v=-1/300,$
$\cot\theta=1/100,$ then there is exactly one bound state at $k=
i\beta_1,$ with
$\beta_1=0.043812\overline{3}$
and the Gel'fand-Levitan norming constant
 $g_1=0.2407\overline{8},$ and no eligible resonances. In other words, in this specific case
we have $N=1$ and $M=1.$ In this case the radius value
given in (4.27) is admissible because it yields $r_1(\ell)>0.$
The corresponding absolute pressure $|P(k,\ell)|$ given in
(2.23) and the corresponding
vocal-tract radius $\overset{\circ}\to r(x)$
given in (4.26) and the
radius $r_1(x)$ given in (4.27) are shown in Figure~6.4.
In the evaluation of $\overset{\circ}\to r(x)$ we have used the Darboux transformation formula
given in (3.7) of [4].

\vskip 10 pt

\centerline{\hbox{\psfig{figure=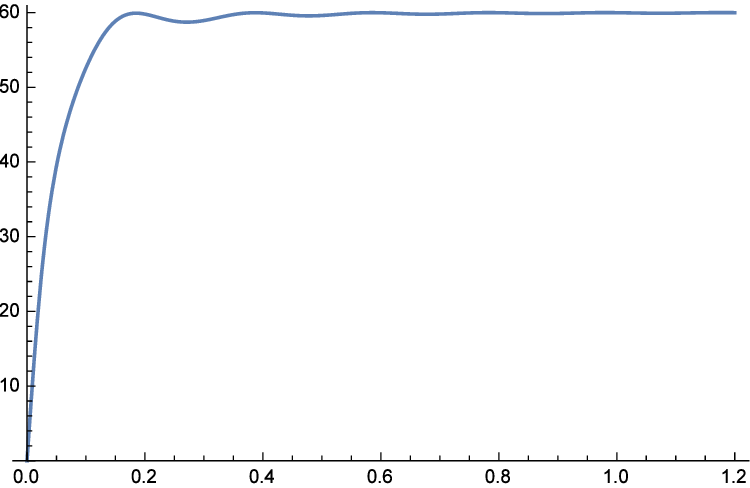,width=2 truein,height=2 truein}\quad
\psfig{figure=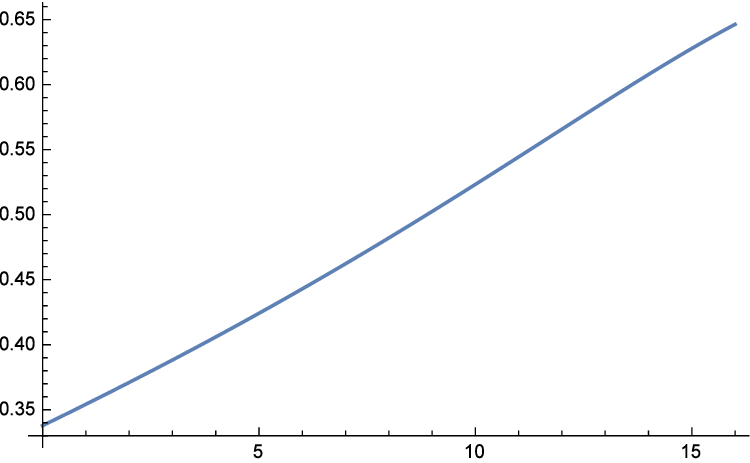,width=2 truein,height=2 truein}\quad
\psfig{figure=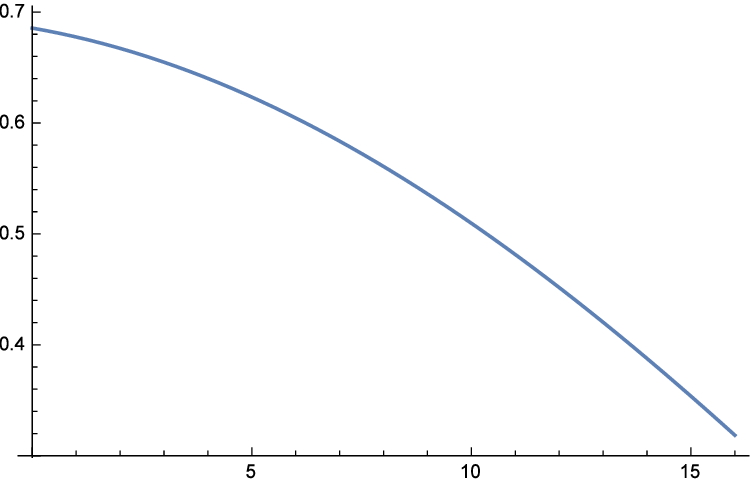,width=2 truein,height=2 truein}}}

\centerline{{\bf Figure~6.4}
$|P(k,\ell)|$ vs $k,$ $\overset{\circ}\to r(x)$ vs $x,$ and $r_1(x)$ vs $x,$
respectively, in Example~6.4(c).}

\vskip 10 pt

\noindent {\bf{REFERENCES}}

\item{[1]} T. Aktosun,
{\it Inverse scattering for vowel articulation with frequency-domain data,}
Inverse Problems {\bf 21}, 899--914 (2005).

\item{[2]} T. Aktosun, {\it
Inverse scattering to determine the shape of a vocal tract,} in: M. A.
Dritschel (ed.), {\it The extended field of operator theory,} Birkh\"auser,
Basel, 2007, pp. 1--16.

\item{[3]} T. Aktosun and J. H. Rose,
{\it Wave focusing on the line,}
J. Math. Phys. {\bf 43}, 3717--3745 (2002).

\item{[4]} T. Aktosun, P. Sacks, and M. Unlu,
{\it Inverse problems for selfadjoint Schr\"odinger operators on the half line with compactly supported potentials,}
J. Math. Phys. {\bf 56}, 022106 (2015).

\item{[5]} T. Aktosun and R. Weder,
{\it Inverse spectral-scattering problem with two sets of
discrete spectra for the radial Schr\"odinger equation,}
Inverse Problems {\bf 22}, 89--114 (2006).

\item{[6]} K. P. Bube and R. Burridge,
{\it The one-dimensional inverse problem of reflection seismology,}
SIAM Rev. {\bf 25}, 497--553 (1983).

\item{[7]} G. Fant, {\it Acoustic theory of speech production,} Mouton,
The Hague, 1970.

\item{[8]} J. L. Flanagan, {\it Speech analysis synthesis and perception,}
2nd ed., Springer, New York, 1972.

\item{[9]} I. M. Gel'fand and B. M. Levitan,
{\it On the determination of a differential equation from its
spectral function,}
Amer. Math. Soc. Transl. (ser. 2) {\bf 1}, 253--304 (1955).

\item{[10]} F. Gesztesy, B. Simon, and G. Teschl,
{\it Zeros of the Wronskian and renormalized oscillation theory,}
Amer. J. Math. {\bf 118}, 571--594 (1996).

\item{[11]} B. M. Levitan, {\it Inverse Sturm-Liouville problems,}
VNU Science Press, Utrecht, 1987.

\item{[12]}  V.\; A.\; Marchenko, {\it Sturm-Liouville operators and
applications,} Birk\-h\"au\-ser, Basel, 1986.

\item{[13]} L. R. Rabiner and R. W. Schafer, {\it Introduction to digital
speech processing,} Now Publ., Hanover, MA, 2007.

\item{[14]} P. E. Sacks,
{\it Reconstruction of step-like potentials,}
Wave Motion {\bf 18}, 21--30 (1993).

\item{[15]} J. Schroeter and M. M. Sondhi, {\it Techniques
for estimating vocal-tract shapes from the speech signal,}
IEEE Trans. Speech Audio Process. {\bf 2}, 133--149 (1994).

\item{[16]} M. M. Sondhi, {\it A survey of the vocal tract inverse problem:
theory, computations and experiments,} in: F. Santosa, Y. H. Pao,
W. W. Symes, and C. Holland (eds.),
{\it Inverse problems of acoustic and elastic waves,} SIAM,
Philadelphia, 1984, pp. 1--19.

\item{[17]} K. N. Stevens, {\it Acoustic phonetics,} MIT Press,
Cambridge, MA, 1998.

\item{[18]} B. H. Story, {\it Comparison of magnetic
resonance imaging-based vocal tract area functions
obtained from the same speaker in 1994 and 2002,}
J. Acoust. Soc. Amer. {\bf 123}, 327--335 (2008).

\item{[19]} W. W. Symes, {\it Impedance profile inversion via
the first transport equation,}
J. Math. Anal. Appl. {\bf 94}, 435--453 (1983).

\item{[20]} J. Weidmann, {\it Linear operators in Hilbert spaces,} Springer-Verlag,
New York, 1980.

\item{[21]} M. Zworski, {\it Distribution of poles for scattering on the real line,}
J. Funct. Anal. {\bf 73}, 277--296 (1987).

\end